\documentclass[journal=jacsat,manuscript=article]{achemso}

\usepackage{comment}
\usepackage[version=3]{mhchem} 
\usepackage{xcolor}
\usepackage{braket}
\usepackage{float}
\usepackage{amsmath}

\newcommand*{\citen}[1]{
 \begingroup
  \romannumeral-`\x 
  \setcitestyle{numbers}
  \cite{#1}
 \endgroup  
}

\author{Nicholas J. Boyer}
\affiliation{Department of Chemistry, University of North Carolina at Chapel Hill, Chapel Hill, North Carolina 27599, USA}
\author{Christopher Shepard}
\affiliation{Department of Chemistry, University of North Carolina at Chapel Hill, Chapel Hill, North Carolina 27599, USA}
\author{Ruiyi Zhou}
\affiliation{Department of Chemistry, University of North Carolina at Chapel Hill, Chapel Hill, North Carolina 27599, USA}
\author{Jianhang Xu}
\affiliation{Department of Chemistry, University of North Carolina at Chapel Hill, Chapel Hill, North Carolina 27599, USA}
\author{Yosuke Kanai}
\affiliation{Department of Chemistry, University of North Carolina at Chapel Hill, Chapel Hill, North Carolina 27599, USA}
\altaffiliation{Department of Physics and Astronomy, University of North Carolina at Chapel Hill, Chapel Hill, North Carolina 27599, USA}
\email{ykanai@unc.edu}

\title[An \textsf{achemso} demo]
 {Machine-Learning Electron Dynamics with Moment Propagation Theory: Application to Optical Absorption Spectrum Computation using Real-Time TDDFT}


\begin{document}

% Yosuke.
\definecolor{ykred}{rgb}{1.0, 0.0, 0.0}
% Ruiyi
\definecolor{rygreen}{rgb}{1.0, 0.5, 0.0}
% Nicholas
\definecolor{nbblue}{rgb}{0.0, 0.0, 1.0}
% Chris
\definecolor{csviolet}{rgb}{1.0,0.0,1.0}
% Jianhang
\definecolor{jxc}{rgb}{0.2,0.8,0.2}
\newcommand{\yk}[1]{{\color{ykred}{#1}}}
\newcommand{\rz}[2]{{\color{rygreen}{#1}}}
\newcommand{\nb}[1]{{\color{nbblue}{#1}}}
\newcommand{\cs}[1]{{\color{csviolet}{#1}}}
\newcommand{\jx}[1]{{\color{jxc}{#1}}}

\begin{abstract}
 We present an application of our new theoretical formulation of quantum dynamics, moment propagation theory (MPT) (Boyer et al., J. Chem. Phys. 160, 064113 (2024)), for employing machine-learning techniques to simulate the quantum dynamics of electrons. In particular, we use real-time time-dependent density functional theory (RT-TDDFT) simulation in the gauge of the maximally localized Wannier functions (MLWFs) for training the MPT equation of motion. Spatially-localized time-dependent MLWFs provide a concise representation that is particularly convenient for the MPT expressed in terms of increasing orders of moments. The equation of motion for these moments can be integrated in time while the analytical expressions are quite involved. In this work, machine-learning techniques were used to train the the second-order time derivatives of the moments using first-principles data from the RT-TDDFT simulation, and this MPT enabled us to perform electron dynamics efficiently. The application to computing optical absorption spectrum for various systems was demonstrated as a proof-of-principles example of this approach. In addition to isolated molecules (water, benzene, and ethene), condensed matter systems (liquid water and crystalline silicon) were studied, and we also explored how the principle of the nearsightedness of electrons can be employed in this context.
\end{abstract}

\section{Introduction}

Real-time simulations of electron dynamics have attracted great interest for studies of non-equilibrium behavior in molecular systems.\cite{doi:10.1063/5.0057587, DRAEGER2017205, doi:10.1021/jacs.3c08226, doi:10.1021/acs.jpcb.3c05446} In particular, real-time time-dependent density functional theory (RT-TDDFT) has become a widely used tool to investigate various phenomena such as optical absorbance \cite{PhysRevLett.127.077401, 10.1063/5.0066753, doi:10.1021/acs.jpcc.9b00296}, energy transfer \cite{doi:10.1021/acs.jctc.2c00600, D0CP04206D}, plasmons \cite{doi:10.1021/acsnanoscienceau.2c00061, Ma2015}, charge transfer \cite{doi:10.1021/acs.jpcb.0c11489,https://doi.org/10.1002/qua.25096}, electronic circular dichroism \cite{10.1063/5.0038904}, thermalization \cite{PhysRevB.93.024306}, high harmonic generation \cite{doi:10.1021/acs.jpca.3c07865, Okyay2022}, electronic stopping \cite{PhysRevLett.130.118401}, electrical conductivity \cite{PhysRevB.107.115131}, photocatalysis \cite{YAMIJALA2022127026}, transient absorption spectroscopy \cite{doi:10.1021/acs.jpclett.2c03599}, spin transfer \cite{doi:10.1021/acs.nanolett.1c00520}, magnons \cite{doi:10.1021/acs.jctc.9b01064}, core electron excitation \cite{PhysRevLett.123.066401}, exciton dynamics \cite{doi:10.1021/acs.jpcc.5b00263}, laser-induced water splitting \cite{PhysRevB.96.115451}, and many other electronic excitation phenomena. The approach of RT-TDDFT is to propagate single-particle time-dependent Kohn Sham (TD-KS) orbitals to study quantum dynamics.\cite{doi:10.1063/5.0057587} These orbitals posses a gauge freedom where a unitary transformation has no effect on the quantum dynamics \cite{JIA201921, 10.1063/1.5095631}. One example of this gauge is the maximally localized Wannier functions (MLWFs) where the orbitals are unitary transformed in spatially-localized orbitals \cite{10.1063/1.5095631}. These MLWFs have previously been found to be useful in studying novel phenomena in complex systems such as Floquet engineering \cite{doi:10.1021/acs.jpclett.1c01037} and the electronic stopping response in DNA \cite{PhysRevLett.130.118401}.
These MLWFs also have been used in RT-TDDFT for efficient implementation of hybrid exchange-correlation (XC) functionals\cite{chris_hybrid}.
Performing RT-TDDFT simulations, however, requires a large computational cost especially for simulating condensed matter systems \cite{doi:10.1063/5.0057587}. 

In recent years, molecular dynamics (MD) simulations have employed machine learning (ML) techniques to speed up calculations \cite{10.1063/5.0047760}. By either predicting the force or the potential energy from atomic positions, the MD-ML models can provide highly accurate simulations with lower computational cost as compared to first-principles MD \cite{doi:10.1021/acs.jctc.0c01343, Han_2018}. This has motivated investigations into using these ML techniques for electron dynamics \cite{Schiffer_ANN1, Schiffer_ANN2}. Secor et al., for example, proposed using artificial neural networks (ANN) as propagators in quantum dynamics for predicting the one-dimensional wavefunction at a future time step using the current time dependent wavefunction and potential \cite{Schiffer_ANN2}. However, training and choosing basis sets for higher dimensional systems proved challenging \cite{Schiffer_ANN2}. 

In our previous work, we proposed a novel theoretical formulation for quantum dynamics in the single-particle description\cite{Boyer_2024}. Our new approach, moment propagation theory (MPT), describes a single-particle wavefunction in terms of increasing orders of moments. We analytically derived the equation of motion for these moments. The proof-of-principle simulations employed up to the fourth-order of moments and accurately modeled the quantum dynamics of both harmonic and anharmonic well systems. 
Motivated by the analytical solution for the harmonic well, the work also proposed using ML techniques to circumvent the expensive calculation of the the moment time-derivatives. An artificial neural network (ANN) model accurately simulated the harmonic potential with low computational cost \cite{Boyer_2024}. This is analogous to the MD-ML models that calculate the force on the atoms through approaches like ANN models. 

In this work, we demonstrate the use of the moment propagation theory with machine-learning techniques (MPT-ML) for real systems through the use of RT-TDDFT simulation.
By using the moments of the spatially localized time-dependent MLWFs, only the moments up to a low order are needed to concisely describe the system within the MPT framework. We demonstrate the accuracy of the MPT-ML model approach for single molecular systems of water, benzene, and ethene. As a test of performance, we compute the optical absorption spectra for these molecules. We then investigate its application to liquid water and crystalline silicon and also examine how the principle of the nearsightedness of electrons can be utilized.

\section{Theoretical and Modeling Details}

\subsection{Brief Overview of Moment Propagation Theory}

In our earlier work \cite{Boyer_2024}, we showed that the single-particle quantum dynamics can be formulated in terms of the moments of increasing orders instead of propagating the wave function using via a Schrodinger-like equation, as done using TDKS equations in RT-TDDFT simulation.
Let us briefly summarize the key points of this MPT that are relevant in this work. 
We express the single-particle wavefunction here as $\psi(x, y, z, t)$ as the orders of the moments are generally not the same in the three Cartesian coordinates. 
The moments of the single-particle probability density is

\begin{equation}
\braket{x^ay^bz^c}(t)= \int \int \int {x^ay^bz^c n(x, y, z,t)d x d y d z},
\end{equation}
where $a,b,c$ are non-negative integers used to denote the a-th, b-th, c-th moment in $x,y,z$ directions of the Cartesian coordinate space and $n(x, y, z, t)$ is the single particle (probability) density, which is the square modulus of the single-particle wave function (i.e. $n(x, y, z,t)=| \psi(x, y, z, t)|^2$). 
The explicit equation of motion for the moments can be derived from the TDSE where 
the first-order time derivative of the moments is 
\begin{equation}
\begin{aligned}
\label{eq:firstder}
\frac{d \braket{x^ay^bz^c}(t)}{d t} 
= & -\frac{i}{2} \int \left[\nabla^2\left(x^ay^bz^c\right)n + 2\nabla\left(x^ay^bz^c\right)\cdot\nabla\psi\psi^*\right]d^3r,\\
\end{aligned}
\end{equation}
and the second-order time derivative of the moments is 
\begin{equation}
\begin{aligned}
\label{eq:secondder}
\frac{d^2 \braket{x^ay^bz^c}(t)}{d t^2} = & \int \operatorname{Re} \left[ - \nabla (x^ay^bz^c) \cdot \nabla V n + \frac{1}{4} \nabla^4\left(x^ay^bz^c\right)n - ( \nabla \otimes \nabla \left(x^ay^bz^c\right) \cdot (\nabla \otimes \nabla \psi)) \psi^* \right] d^3r.\\
\end{aligned}
\end{equation}

In our earlier work\cite{Boyer_2024}, we showed that the numerical quantum dynamics simulation scheme can be formulated by Taylor-expanding the moments in time and truncating the expansion at the second order, as done in most classical molecular dynamics simulation. 
Importantly, we showed that the second-order time derivative can be given in terms of the moments and their first-order time derivatives albeit the explicit expression is highly complicated for numerical evaluation, especially for higher-order moments. 

\subsection{Machine Learning the second-order time derivatives}
In general, the second-order time derivative can be expressed as
\begin{equation}
\label{eq:eom_form}
\frac{d^2 \braket{x^ay^bz^c}(t)}{d t^2} = F ( 
\{\braket{x^dy^ez^f}(t)\},
\{\frac{d \braket{x^dy^ez^f}(t)}{dt}\}),
\end{equation}
where $F$ is a function, generally very complicated, of the moments and their first-order time derivatives. 
$F$ can be solved analytically only for very limited cases such as the harmonic oscillator as discussed in our earlier work \cite{Boyer_2024}. 
We also proposed that the use of ML approaches including a simple ANN model for $F$ instead of its explicit evaluation, as often done for potential energy in classical MD simulation \cite{Boyer_2024}. 
 While it is tempting to apply popular ML techniques like ANN and deep-learning models, it is also possible to use more traditional ML techniques by incorporating known physical behavior.
For example, Vulpe and coworkers developed a MD potential using the physics-based many-body expansion of the potential energy \cite{doi:10.1021/acs.jctc.2c00645}. 
Hauge and coworkers noted that ANNs struggle to extrapolate data, such as dipole moments in time, so they had to enforce certain restrictions to prevent over-fitting and ensure stable extrapolation \cite{doi:10.1021/acs.jctc.3c00727}.
In this work, with the analytical expression (Eq. \ref{eq:eom_form}) in mind, 
we examine the linear model for the ML,

\begin{equation}
\frac{d^2 \braket{x^ay^bz^c}(t)}{d t^2} = B_{a,b,c} + \sum_{d,e,f} C_{a,b,c,d,e,f} \braket{x^dy^be^f}(t)+ \sum_{d,e,f} D_{a,b,c,d,e,f}\frac{d \braket{x^dy^ez^f}(t)}{d t},
\end{equation}
where the coefficients $B,C,D$ are to be machine-learned from RT-TDDFT simulation. 
For brevity, let us denote the moments using $\braket{\mathbf{r}^i}(t) \equiv \braket{x^ay^bz^c}(t)$. For multi-electron systems in the TD-KS scheme, interactions between electrons must be incorporated as well

\begin{equation}
\label{eq:linearmodel}
\frac{d^2 \braket{\mathbf{r}^i}_j}{d t^2} = B_{i, j} + \sum_{k, l} C_{i, j,k,l} \braket{\mathbf{r}^l}_k + \sum_{k, l} D_{i, j,k,l}\frac{d \braket{\mathbf{r}^l}_k}{d t},
\end{equation}
where $\braket{\mathbf{r}^i}_j$ is the i-th moment of the j-th electronic state.

\subsection{Time-Dependent Kohn Sham Equations in Wannier Gauge}

In extended periodic systems, the Bloch states satisfy
$\psi_n(\mathbf{r}+\mathbf{R}, \mathbf{k}) = e^{i\mathbf{k} \cdot \mathbf{R}} \psi_n(\mathbf{r}, \mathbf{k})$ where $n$ is the band index and $\mathbf{R}$ is the lattice-periodic cell vector. Correspondingly, Wannier functions are given by

\begin{equation}
w_n(\mathbf{r}, \mathbf{R}) = \frac{\Omega}{(2 \pi)^3} \int_{BZ} d\mathbf{k} e^{-i\mathbf{k} \cdot \mathbf{R}} \psi_n(\mathbf{r}, \mathbf{k}),
\end{equation}
where $\Omega$ is the volume of the real-space periodic cell \cite{10.1063/1.5095631}. Wannier functions are translationally invariant such that $w_n(\mathbf{r}, \mathbf{R}) = w_n(\mathbf{r}- \mathbf{R}, \mathbf{0})$, and it can be denoted simply as $w_n(\mathbf{r})$. These Wannier functions possess a gauge freedom, and it has become popular to make these Wannier functions unique by minimizing the total spread given as

\begin{equation}
S = \sum_l (\braket{w_l | \mathbf{r}^2 | w_l} - \braket{w_l | \mathbf{\hat{r}} | w_l}^2),
\end{equation}
where the position operator $\hat{\mathbf{r}}$ is defined according to Resta's formula in extended systems \cite{PhysRevLett.80.1800},

\begin{equation}
\braket{\mathbf{\hat{r}}} = \frac{\mathbf{L}}{2\pi} \operatorname{Im} \ln{\braket{e^{i\frac{2 \pi}{\mathbf{L}} \cdot \mathbf{\hat{r}}}}},
\end{equation}
where $\mathbf{L}$ is the lattice vector of the periodic cell. 
This can be extended to second order of moments as well\cite{thesismoments}

\begin{equation}
\braket{(\hat{r} - \braket{\hat{r}})^2} = ( \frac{L_r}{2\pi} )^2 (1 - |\braket{e^{i\frac{2 \pi}{L_r} \hat{r}}}|^2),
\end{equation}
where r is one of the Cartesian coordinates $x$, $y$, or $z$ and

\begin{equation}
\braket{ (\hat{r} - \braket{\hat{r}}) ( \hat{r}' - \braket{\hat{r}'} )} = \frac{L_r L_{r'}}{16\pi^2} \left(\ln{|\braket{e^{i\frac{2 \pi}{L_r} \hat{r}} e^{-i\frac{2 \pi}{L_{r'}} \hat{r}'}}|^2} -\ln{|\braket{e^{i\frac{2 \pi}{L_r} \hat{r}} e^{i\frac{2 \pi}{L_{r'}} \hat{r}'}}|^2}\right),
\end{equation}
where r is one of the Cartesian coordinates $x$, $y$, or $z$ and $r'$ is another Cartesian coordinate.

As discussed in Ref\citen{10.1063/1.5095631}, TD-KS equations can be propagated using the
MLWFs, 
\begin{equation}
i\frac{\partial w_l(\mathbf{r}, t)}{\partial t} = 
\{
\hat{U}^{ML} 
-\frac{1}{2}\nabla^2 + V_{KS}(\mathbf{r}, t)
\}
w_l(\mathbf{r}, t),
\end{equation}
where $\hat{U}^{ML}$ is the operator for ensuring the maximally localization of the Wannier functions, and $V_{KS}$ is the KS potential. This scheme has been successfully used for various applications and also for reducing the computational cost of evaluating the exact exchange \cite{chris_hybrid}. 

\subsection{Computational Method}

Instead of using the Taylor series expansion as often done in classical MD simulation (see Supporting Information), we propose an alternative scheme.
By applying the MPT to the quantum dynamics described by TD-MLWFs, solutions to the linear model (LM) (Eq. \ref{eq:linearmodel}) can be obtained analytically. 
Eq. \ref{eq:linearmodel} is written in terms of matrices as

\begin{equation}
\label{eq:matrix_eq}
\Ddot{\mathbf{X }}(t) = \mathbf{C} \mathbf{X }(t) + \mathbf{D} \Dot{\mathbf{X }}(t) + \mathbf{B},
\end{equation}
where $\mathbf{X}(t)$ is the vector with the moments,
$\mathbf{X }(t) = (<\mathbf{r}^{i=1}>_{j=1}(t), <\mathbf{r}^{i=1}>_{j=2}(t), ...,<\mathbf{r}^{i=2}>_{j=1}(t),... )^T
$, 
$\mathbf{B}$, $\mathbf{C}$, and $\mathbf{D}$ are specified in Eq. \ref{eq:linearmodel}.
It is convenient to rewrite this equation as
\begin{equation}
\Dot{\mathbf{Y}}(t) = \mathbf{A} \mathbf{Y}(t) + \mathbf{E},
\end{equation}
where we define
\begin{equation}
\mathbf{Y}(t) \equiv \begin{bmatrix}
\mathbf{X}(t) \\
\Dot{\mathbf{X}}(t)
\end{bmatrix},
\end{equation}

\begin{equation}
\mathbf{A} \equiv \begin{bmatrix}
\mathbf{0} & \mathbf{I} \\
\mathbf{C} & \mathbf{D} \\
\end{bmatrix},
\end{equation}

\begin{equation}
\mathbf{E} \equiv \begin{bmatrix}
\mathbf{0} \\
\mathbf{B}
\end{bmatrix}.
\end{equation}
The matrices $\mathbf{0}$ and $\mathbf{I}$ are zero and identity matrices respectively with the same size as other matrices.
The general solution to this linear ordinary differential equation (ODE) can be expressed as
\begin{equation}
\mathbf{Y}(t)= e^{\mathbf{A}t}\mathbf{V} - \mathbf{A}^{-1}\mathbf{E},
\end{equation}
where $\mathbf{V}$ is defined by
\begin{equation}
\mathbf{V} \equiv \mathbf{Y}(0) + \mathbf{A}^{-1}\mathbf{E}.
\end{equation}
The initial value problem (IVP) here is 
\begin{equation}
\label{eq:mainEQ}
\mathbf{Y}(t) = \mathbf{P} e^{\mathbf{Q} t} \mathbf{P}^{-1} \mathbf{V} - \mathbf{A}^{-1}\mathbf{E},
\end{equation}
where $\mathbf{Q}$ is the diagonal matrix such that $\mathbf{A} = \mathbf{P} \mathbf{Q} \mathbf{P}^{-1}$.
$\mathbf{Y}(t)$ contains time-dependent information about the moments. When the Fourier transform of the first-order moments in $\mathbf{Y}(t)$ are used to calculate the optical absorption spectrum as discussed in the following, the eigenvalues ($\mathbf{Q}$) can be identified as the frequencies of the absorption spectrum and the eigenvectors ($\mathbf{P}$) provide the magnitude of the transition dipoles. 

In few cases, the matrix $\mathbf{Q}$ has diagonal elements with $\operatorname{Re}\mathbf{Q}_{ii} > 0$ such that the solution diverges in the form of $e^{\operatorname{Re}\mathbf{Q}_{ii}t}$. 
These eigenvalues tend to be close to zero, but having such real-valued non-zero elements lead to an nonphysical solution for MPT in numerical simulations. We correct this numerical artifact by setting the real part of these eigenvalues to zero to eliminate the diverging solution. Generally speaking, the cause of having positive real eigenvalues stems from fitting the linear model to a data set produced by RT-TDDFT simulation with a finite simulation time $T$. Indeed, we observe that as $T$ increases, the need for this correction decreases. 
In numerically performing some of these matrix operations, we may use other standard corrections. 
In evaluating $\mathbf{A}^{-1} = \mathbf{P}\mathbf{Q}^{-1}\mathbf{P}^{-1}$, we set any eigenvalues with their absolute value below a certain threshold $ |\mathbf{Q}_{ii}| < c$ (we use c=0.005) as $\mathbf{Q}_{ii}^{-1} = 0$. Likewise, we set $\mathbf{Q}_{ii}^{-1} = \frac{1}{\mathbf{Q}_{ii}}$ when $ |\mathbf{Q}_{ii}| \ge c$.
Additionally, we employ a high eigenvalue cutoff $h$ such that $(\mathbf{P}^{-1}\mathbf{V})_i = 0$ if $|\mathbf{Im} (\mathbf{Q}_{ii})| > h$ (we typically use h=2 or 54.4 eV), removing nonphysical high frequency noises.

\noindent
\textbf{Ridge Regression:}
Regularization refers to a statistical technique to minimize errors from overfitting with training data, and so-called ridge regression is one of the most commonly-employed regularization technique for linear regression models. 
With a large number of variables as for condensed matter systems, the overfitting becomes a practical issue because of the multicollinearity within the dynamics of MLWFs. Thus, we also examined the effectiveness of the ridge regression technique, which minimizes the loss function

\begin{equation}
L = \alpha \sum_i M_i^2 + \sum_j (y_j - LM(x_j))^2,
\end{equation}
where $LM$ is the linear model as described above in Eq. \ref{eq:matrix_eq}.
The $M_i$ is the i-th learnable parameter in matrix $\mathbf{A}$.
The variables $x_j$ and $y_j$ are the $\mathbf{Y}(t_j)$ inputs and the $\mathbf{\Ddot{\mathbf{X}}}(t_j)$ outputs, respectively where $j$ is for the time index. 
The hyperparameter, $\alpha$, is an additional adjustable parameter that is used to reproduce the training data closely.

\noindent
\textbf{Nearsightedness of Electrons:} 
With increasingly large numbers of the moments for modeling condensed matter systems, numerical noises from fitting the first-principles data could degrade the accuracy.
The nearsightedness principle of electrons \cite{PhysRevLett.76.3168, Nearsight_2} can be invoked to reduce the number of the parameters necessary in the above proposed model based on the moment propagation theory.
According to the nearsightedness principle, 
local electronic properties like the probability density depend on the effective external potential of only nearby points. Changes in that potential beyond a certain distance have limited effects on the local properties.
This allows us to introduce a cutoff distance beyond which the electrons (represented by MLWFs) are not impacted the dynamics of other electrons. 
Then, the equation of the motion for the moments (Eq. \ref{eq:linearmodel}) can be written in terms of the subset of the all MLWFs as
\begin{equation}
\frac{d^2 \braket{\mathbf{r}^i}_j}{d t^2} = B_{i, j} + \sum_{k, l\in |\braket{\mathbf{r}^i}_j - \braket{\mathbf{r}^l}_k| < r_c} C_{i, j,k,l} \braket{\mathbf{r}^l}_k + \sum_{k, l\in |\braket{\mathbf{r}^i}_j - \braket{\mathbf{r}^l}_k| < r_c} D_{i, j,k,l}\frac{d \braket{\mathbf{r}^l}_k}{d t},
\end{equation}
where $r_c$ is the cutoff distance beyond which the MLWFs do not impact their dynamics. 
In addition to allowing us to develop an effective computational method, this procedure also enable us to study the extent to which the nearsightedness principle applies in real systems. 
Studying the necessary cutoff distance for fully reproducing the RT-TDDFT result informs us about the effective distance for the such nearsightedness of electrons in condensed matter systems.

\begin{figure}[H]
  \includegraphics[ width=0.9\linewidth]{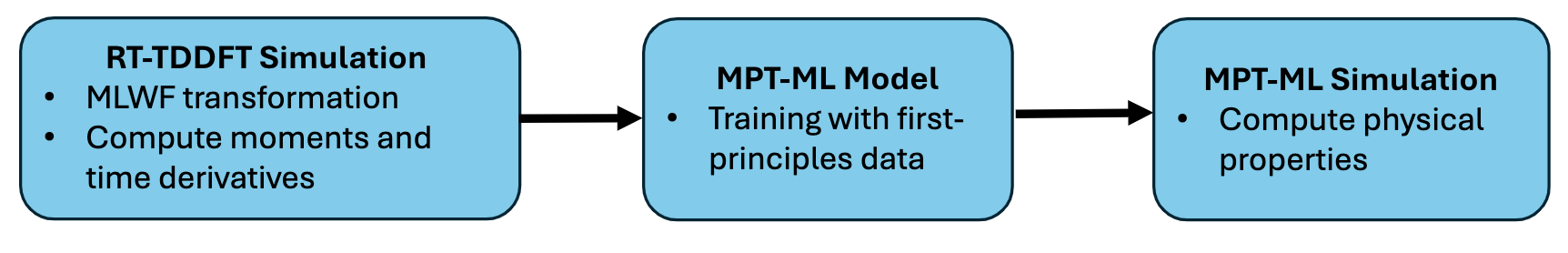}

  \caption{Workflow diagram for the MPT-ML model. First, the RT-TDDFT simulation is performed, and the moments and their time derivatives are computed for the time-dependent MLWFs. Then, the machine-learning model is developed by training the equation-of-motion of the moment propagation theory (MPT) using this first-principles data. Finally, the resulting MPT-ML model is used to simulate electron dynamics to compute physical properties such as optical absorption spectrum.} 
  \label{fig:work}
\end{figure}

The workflow of this work is summarized in Figure \ref{fig:work}. First, the RT-TDDFT simulation is performed using the Qb@ll code\cite{10.1063/1.5095631}. In the RT-TDDFT simulation with the MLWF gauge, all the moments are computed at each time step. 
The machine-learning model is then developed by fitting the equation-of-motion from the moment propagation theory (MPT-ML) to this first-principles training data. 
The resulting MPT-ML model is examined against the RT-TDDFT simulation by computing the optical absorption spectra, which contain the electronic excitation at all frequencies. 

% Results
\section{Results and Discussion}
\subsection{Calculation of Dielectric Function}

To demonstrate the above described approach based on the moment propagation theory in the context of RT-TDDFT, optical absorption spectra are calculated. 
For extended systems, the optical absorption spectrum can be obtained from the imaginary part of the dielectric function \cite{10.1063/1.5095631}, 
\begin{equation}
\label{eq:fourier}
\epsilon(\omega) = 1 + \frac{4\pi }{3 } \operatorname{Tr} \left[ \sigma_{\mu \nu}(\omega) \right],
\end{equation}
where $\sigma_{\mu \nu}(\omega)$ is the complex frequency-dependent polarizability tensor. It can be obtained by Fourier transforming the time-dependent polarization
\begin{equation}
\label{eq:ft}
\sigma_{\mu \nu}(\omega) = \frac{1}{\delta_{\nu}} \int dt e^{i \omega t} \sum_j
\braket{r_{\mu}}_j (t),
\end{equation}
where $\braket{r_{\mu}}_j (t)$ is the first-order moment that is propagated as vector elements of $\mathbf{X}(t)$ (Eq. \ref{eq:mainEQ}).
Here $\delta_\nu$ is the strength of the abrupt homogeneous electric field applied to the system in $\nu$ direction using the length gauge \cite{10.1063/1.5095631}. 
The imaginary part of the dielectric function is directly related to the optical absorbance while the real part is related to the dispersion. For isolated systems such as gas-phase molecules the macroscopic dielectric function is not well defined, and the optical absorption is typically described in terms of the dipole strength function, which is also expressed in terms of the polarizability tensor as
\begin{equation}
\label{eq:sw}
S(\omega) = \frac{4 \pi \omega}{3c} \operatorname{Tr} \left[\operatorname{Im} \sigma_{\mu \nu}(\omega) \right],
\end{equation}
where $c$ is the speed of light.
In practice, we add a damping term in the form of $e^{-\frac{t}{\tau}}$ in Eq. \ref{eq:ft} where $\tau$ is chosen to be $\sim$100 a.u.. This damping term reduces the noise from having the finite amount of dynamics in taking the Fourier transform.

\subsection{Optical absorption spectrum of gas phase molecules}

To investigate the applicability of the above-described approach of using the machine learning linear model for the moment propagation theory (MPT-ML) approach in practice, we consider several isolated molecules of water, benzene, and ethene. 
For RT-TDDFT simulation, the PBE XC functional \cite{PBE} was used with 40 Rydberg cutoff for planewave expansion and PBE Optimized Norm-Conserving Vanderbilt (ONCV) pseudopotentials were used \cite{SCHLIPF201536}. 
A single molecule is placed in a 70 a.u. cubic simulation cell. A delta kick strength of 0.01 a.u. was used in the applied electric field, and 0.2 a.u. was used for the time step in the enforced time reversal symmetry (ETRS) \cite{ETRS} integrator for a total of 200 a.u. simulation time. 
As discussed above, RT-TDDFT simulation was performed in the Wannier gauge and the individual moments are obtained for each MLWF. A key point of this study here is whether the electron dynamics necessary for calculating physical properties like optical absorption spectra can be adequately described using only low orders of the moments. While MPT is exact in principle, its practical advantage is limited by the orders of the moments necessary for describing electron dynamics in real systems.

\begin{figure}[H]
  \includegraphics[clip, width=0.95\linewidth]{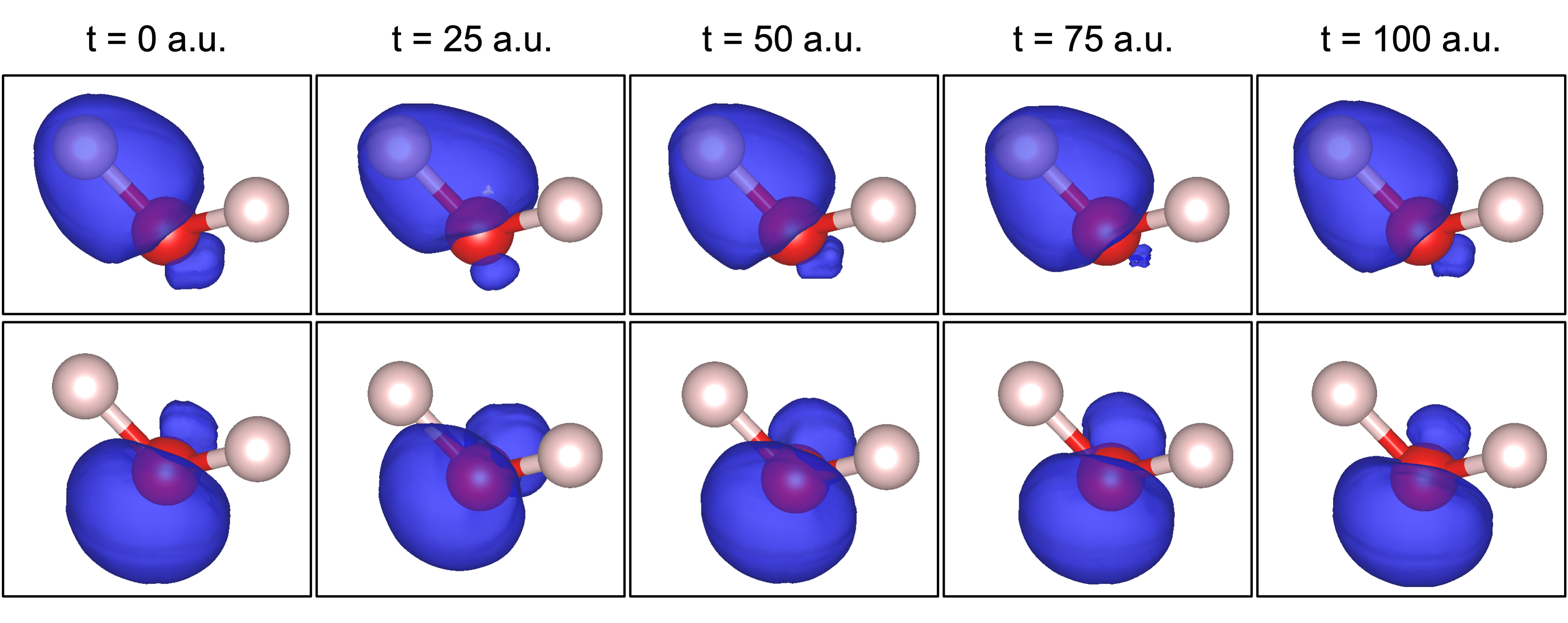}
  \caption{Isosurfaces of a single O-H bond-centered TD-MLWF (top panels) and an oxygen lone pair TD-MLWF (bottom panels) for an isolated water molecule. Each snapshot captures the TD-MLWFs at different instances of time during an RT-TDDFT simulation, following perturbation by an impulsive electric field.} 
  \label{fig:watermlwf}
\end{figure}

Figure \ref{fig:watermlwf} shows the dynamics of a MLWF on a single water molecule in RT-TDDFT.
The MPT-ML approach seeks to capture the dynamics of this single MLWF using increasing orders of moments. As figure \ref{fig:watermlwf} shows, the MLWF is highly localized and amenable to using a concise description using low orders of moments. This remains the case as the MLWF changes over time allowing the use of ML methods to learn the dynamics of moments in the MPT framework.

\begin{figure}[H]
  \includegraphics[ width=0.9\linewidth]{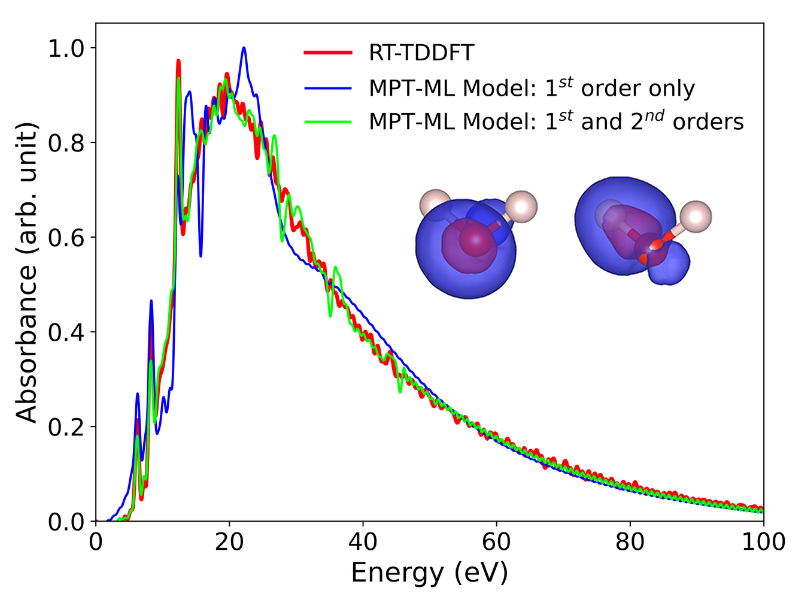} 

  \caption{Optical absorption spectra of a single water molecule, obtained using the RT-TDDFT simulation and MPT-ML model. The MPT-ML model was performed with up to the second-order moments. The contours of the electron density for the O-H bond and oxygen lone pair MLWFs are shown. See Figure S2 in Supporting Information for a close-up figure focused on the energy range below the broad peak. }
  \label{fig:waterabs}
\end{figure}

Figure \ref{fig:waterabs} shows that the results from the MPT-ML approach and the reference RT-TDDFT result, which also serves as the training dataset. 
The MPT-ML approach uses up to the second-order moments and their time derivatives in Eq. \ref{eq:mainEQ}. The optical absorption spectrum show three prominent sharp peaks of 6.2 eV, 8.3 eV, and 12.4 eV below the broad peak centered at 20 eV. With the first-order moments only, the MPT-ML model captures the first two peaks at 6.2 eV and 8.3 eV well but it fails to reproduce the third peak at 12.4 eV. 
By including up to the second-order moments, the MPT-MP model is able to correctly capture also the third peak in addition to the rest of the spectrum features. By using a more complete description of the MLWFs of the single water system with higher orders of moments the result is expected to match the RT-TDDFT result. We also notice that since the size of matrix $\mathbf{Q}$ from the IVP is larger for the second order moments, there are more frequencies that could exist in $\mathbf{Q}$. This is seen as the increasing roughness of the second order result over the first order. 

\begin{figure}[H]
  \includegraphics[width=0.9\linewidth]{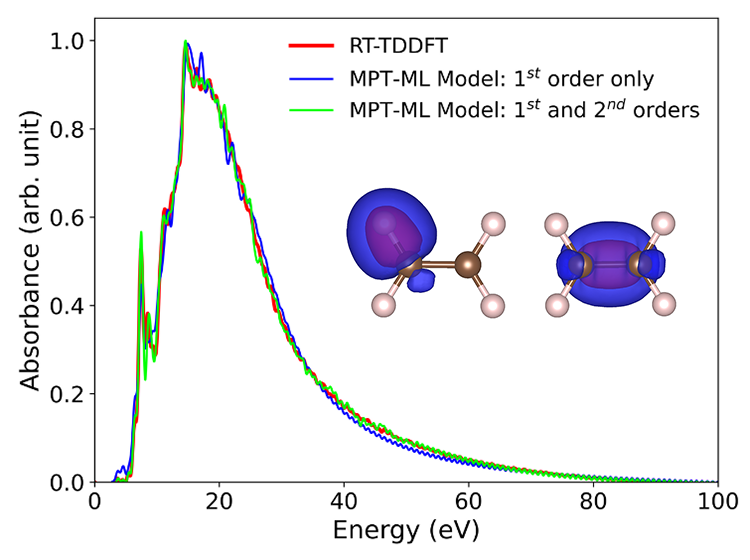}

  \caption{Optical absorbance spectra of an ethene molecule, obtained using the RT-TDDFT simulation and MPT-ML model. The MPT-ML model was performed with up to the second-order moments. The contours of the electron density for the C-H bond and C=C bond MLWFs are shown. See Figure S3 in Supporting Information for a close-up figure focused on the energy range below the broad peak. }
  \label{fig:ethene}
\end{figure}

% Ethene
We apply the MPT-ML approach here on an ethene molecule to examine its applicability for molecules with double bonds. The optical absorption spectrum show a single sharp peak at 7.5 eV below the broad peak centered at 20 eV as seen in Figure \ref{fig:ethene}. In this case, the MPT-ML model well reproduces the spectrum even with the first-order moments only, and 
including up to the second-order moments only further make the spectrum better as in the case of RT-TDDFT result.

\begin{figure}[H]
  \includegraphics[ width=0.9\linewidth]{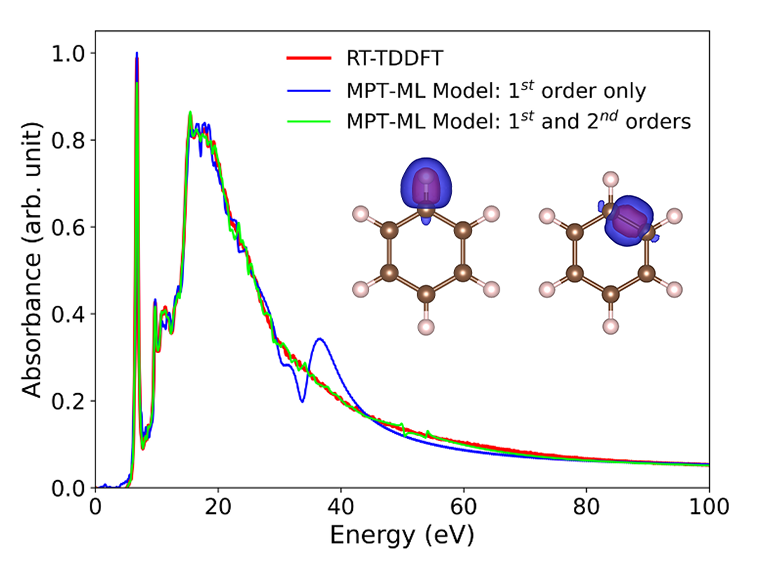}

  \caption{Optical absorption spectra of a benzene molecule obtained using the RT-TDDFT simulation and MPT-ML model. The MPT-ML model was performed with up to the second-order moments. The contours of the electron density for the C-H bond and C=C bond MLWFs are shown. See Figure S4 in Supporting Information for a close-up figure focused on the energy range below the broad peak. }
  \label{fig:benzene}
\end{figure}

% Benzene 
A benzene molecule was studied here particularly because of the delocalized nature of electrons as manifested in conjugation around the carbon atoms. The same computational parameters were used for RT-TDDFT simulation as in the case of water molecule, except for using a longer simulation time of 400 a.u. 
Figure \ref{fig:benzene} shows the optical absorption spectrum of a single benzene molecule. A notable feature is the prominent absorption peak at 6.8 eV, and this key feature is accurately reproduced by the MPT-ML model.
While the MPT-ML model with only the first-order moments is able to capture this absorption peak correctly, it gives an erroneous broad peak at 40 eV. 
By including up to the second-order moments, the MPT-ML is able to correctly eliminate this behavior, yielding an accurate absorption spectrum. 

% Liquid water and silicon
\subsection{Optical absorption spectrum of condensed matter systems}
We examine here the MPT-ML approach for more complex systems of condensed matter systems. In particular, we consider the case of liquid water and crystalline silicon.

\noindent
\textbf{Liquid Water:}
For liquid water a cubic simulation cell (30.6683 a.u.) containing 162 water molecules (1296 electrons) with periodic boundaries was used. The structure of liquid water was generated by taking a snapshot of the equilibrated system following a 20 picosecond classical molecular dynamics simulation at 300 K using the single point charge with polarization correction (SPC/E) model \cite{SPC-E}. 
All atoms are held fixed for the RT-TDDFT simulation, and a delta kick strength of 0.01 a.u. with a 0.1 a.u. time step was used by employing the enforced time reversal symmetry (ETRS) \cite{ETRS} integrator for a total of 250 a.u. simulation time. 
The PBE approximation was used for the XC functional, and Hamann-Schluter-Chiang-Vanderbilt (HSCV) pseudopotentials\cite{PhysRevB.32.8412} were used with a 40 rydberg cutoff for the planewave kinetic energy cutoff for the KS orbitals. 
Previous work has shown that this liquid water simulation cell is fully converged with respect to cell size \cite{water2, water1} and that PBE gives an accurate description of the optical absorption spectra \cite{water1}.

\begin{figure}[H]
  \includegraphics[width=0.9\linewidth]{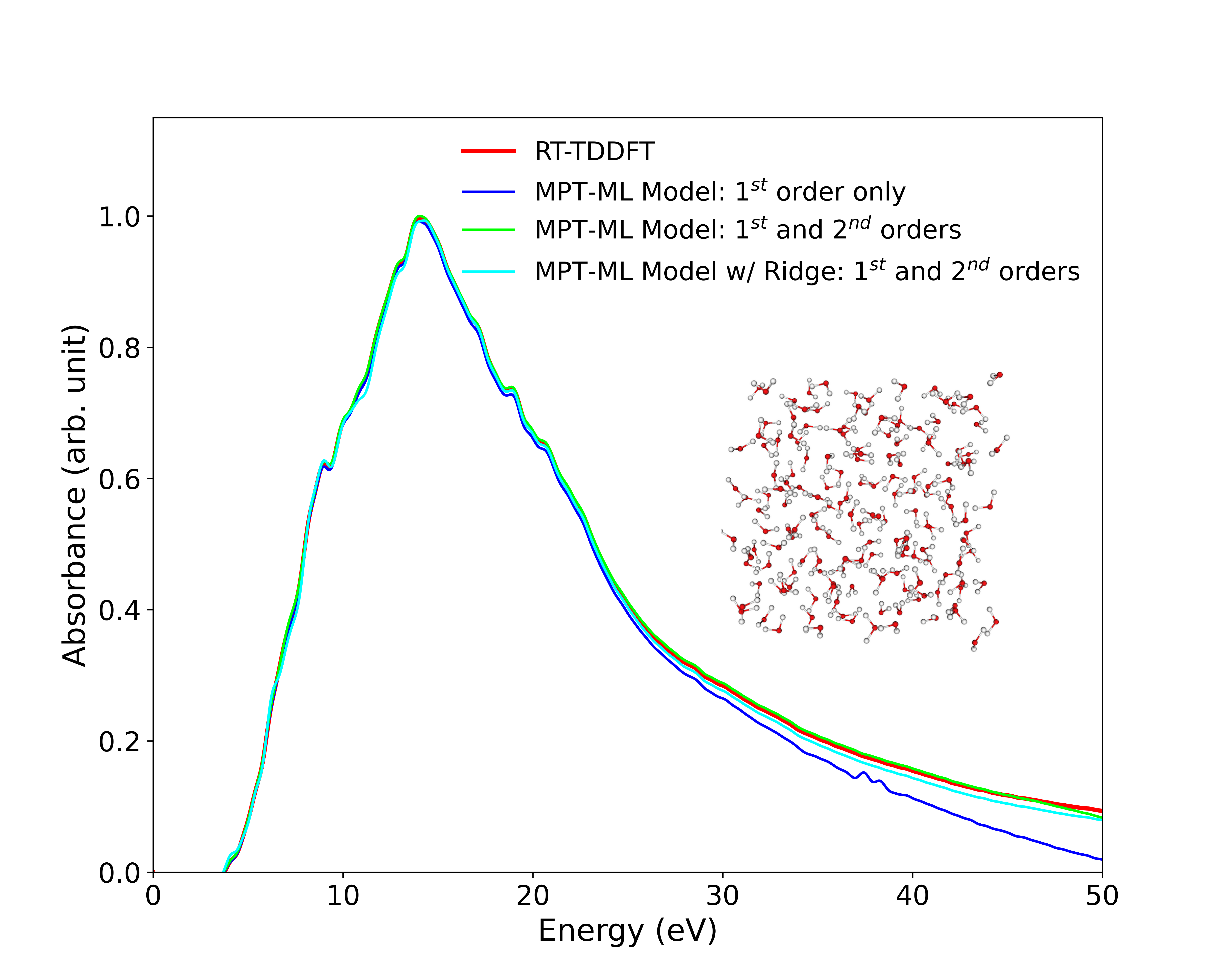}

  \caption{Optical absorption spectra for the 162-molecule liquid water cell using RT-TDDFT simulation and the MPT-ML model. The inset shows the 162 water structure used. The MPT-ML model was employed also with the ridge regression, with the hyperparameter value of $10^{-8}$.}
  \label{fig:waters1}
\end{figure}

In Figure \ref{fig:waters1}, we compare
the MPT-ML model with the RT-TDDFT simulation. As can be seen, by including only the first-order moments in the MPT-ML model already performs quite well in reproducing the RT-TDDFT spectrum. At the same time, the tail end of the spectrum above 30 eV starts to deviate from the first-principles calculation unless the second-order moments are also included. 
For condensed matter systems with a large number of variables for the MPT model, we also examined the use of the ridge regression technique as discussed in the Computational Method section. 
For this particular case of water, the ridge regression does not have much impact unlike the crystalline silicon case discussed in the following section.

\begin{figure}[H]  \includegraphics[width=0.9\linewidth]{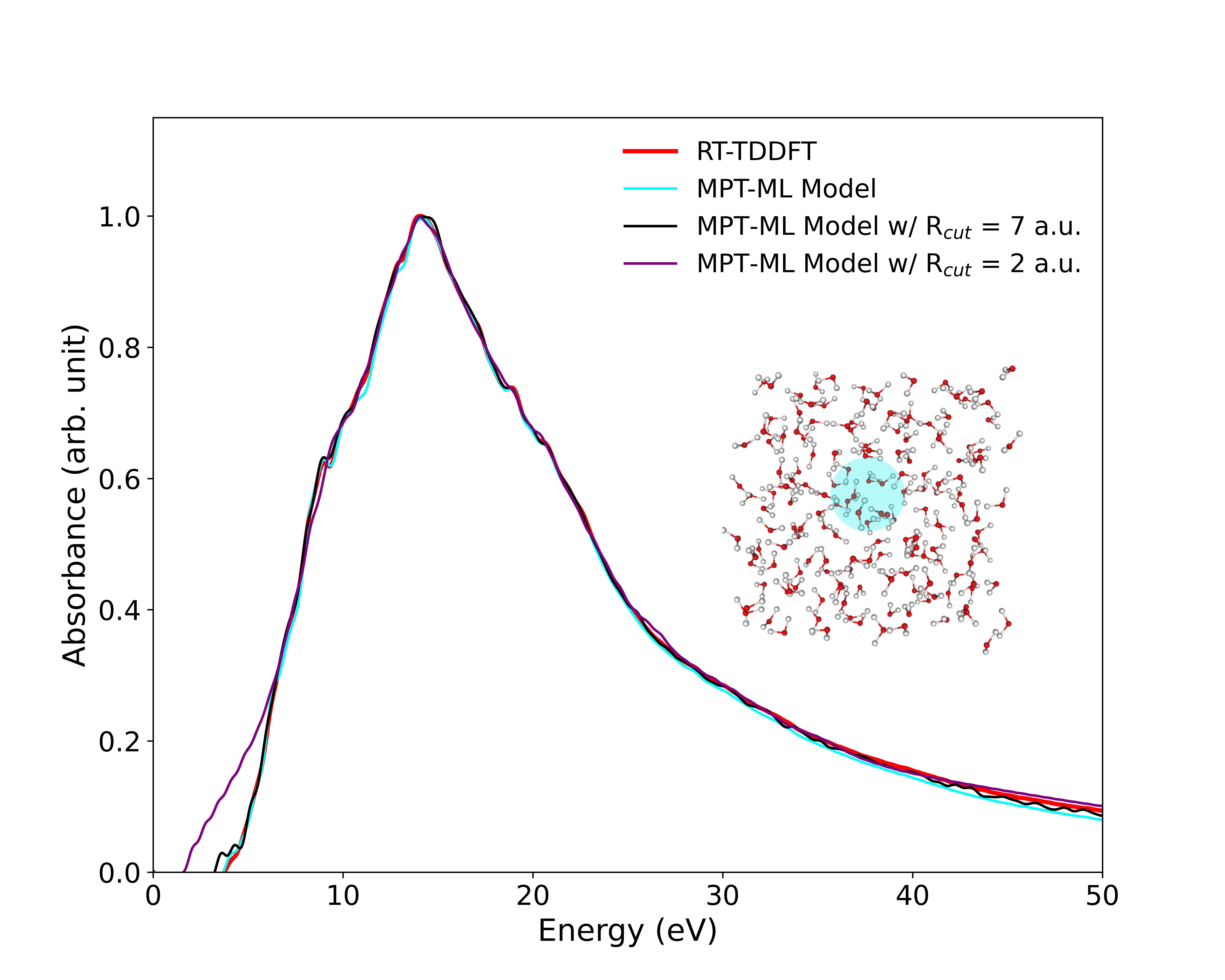}
  
  \caption{Optical absorption spectra for the 162 molecule liquid water cell using RT-TDDFT simulation and the MPT-ML model. The effect of having the varying cutoff radii ($R_{cut}$) on the MPT-ML model is shown. The MPT-ML models use the ridge regression, including the first and second order of moments. The inset shows the 162-molecule liquid water structure, indicating the size of 7 a.u. radius with the blue shade.}
  \label{fig:waters3}
\end{figure}

For linear response properties like the optical absorption spectrum, it is instructive to examine the nearsightedness principle of electrons by Kohn \cite{doi:10.1073/pnas.0505436102} in condensed matter systems.
A particular question in the context of the MPT is to what extent the quantum dynamics of individual Wannier functions can be described by accounting for the dynamics of nearby Wannier functions. 
We examine here such an effective radius of influence for the dynamics of individual Wannier functions, studying the non-local nature of the many-body quantum dynamics for this electronic system.
We do so by introducing the cutoff radius for individual MLWFs in constructing the MPT-ML model as described in the Computational Method section.
Figure \ref{fig:waters3} shows how the optical absorption spectrum changes with the cutoff radius, $R_{cut}$, of 2 and 7 a.u.
The distance of 2 a.u. corresponds to having only the intra-molecular interactions among MLWFs on individual water molecules. 
With the cutoff radius of 7 a.u., the model includes the inter-molecular interactions among MLWFs of their neighboring water molecules. This essentially take into account the dynamical effect within the first solvation shell around individual water molecules. 
The $R_{cut}=$ 7 a.u. spectrum captures all the key features as seen in Figure \ref{fig:waters3} while the $R_{cut}=$ 2 a.u. spectrum shows that it is too short to capture the ``nearsightness" as perhaps expected.
This analysis not only provides valuable insight into the short-range nature of quantum dynamics responsible for the optical absorption in water but also offers an effective scheme to reduce the computational cost of simulating electron dynamics in large complex systems. 

% Table
Importantly in the context of MPT-ML approach, this approach also allows us to significantly reduce the number of parameters to machine-learn. Table \ref{tbl:example} shows the number of moments and the corresponding parameters needed for different systems and settings.
Letting $n$ be the number of moments, the number of parameters to be learned is $\frac{n^2 + n}{2}$. 
In condensed matter systems like water, over 68 million parameters need to be machine-learned even when we need only up to the second-order moments. 
Using $R_{cut}=7 a.u.$, only 4.98\% of these parameters are necessary, significantly reducing the computational complexity of the machine-learning.

\begin{table}[H]
 \caption{Number of parameters in the MPT-ML models}
 \label{tbl:example}
 \begin{tabular}{lllll}
 \hline
System   & Number of Moments & Number of Parameters  \\
\hline
Water   & 72 & 2,628 \\
Benzene  & 270 & 36,585\\
Ethene   & 108  & 5,886 \\
162 Waters & 11,664   & 68,030,280 \\ 
162 Waters ($R_{cut} = 7$ a.u.) & 11,664   & 3,385,476 \\ 
162 Waters ($R_{cut} = 2$ a.u.) & 11,664   & 425,736 \\ 
Silicon & 4,608 & 10,619,136 \\
Silicon ($R_{cut} = 14$ a.u.) & 4,608 & 1,661,184 \\
Silicon ($R_{cut} = 7$ a.u.) & 4,608 & 624,384 \\
\hline
\end{tabular}
\end{table}

\begin{figure}[!h]
  \includegraphics[ width=0.9\linewidth]{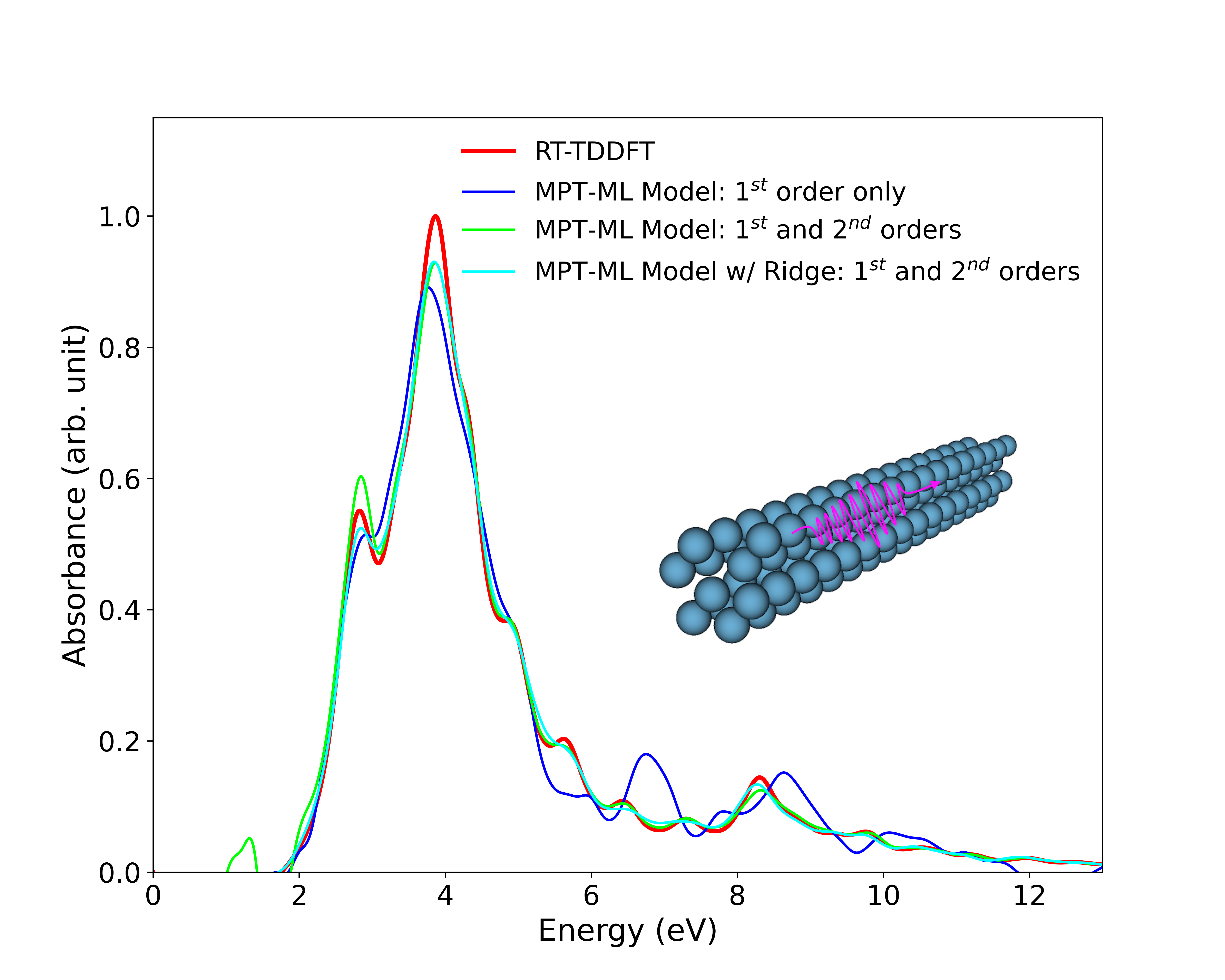}

  \caption{Optical absorption spectra of crystalline silicon, calculated with the 512-electron (128-atom) simulation cell using RT-TDDFT simulation and the MPT-ML model. The MPT-ML model was employed also with the ridge regression, with the hyperparameter value of $10^{-12}$. The inset shows the 128-atom simulation cell used and the direction of the external electric field applied. }
  \label{fig:csi}
\end{figure}

\begin{figure}[!h]
  \includegraphics[ width=0.9\linewidth]{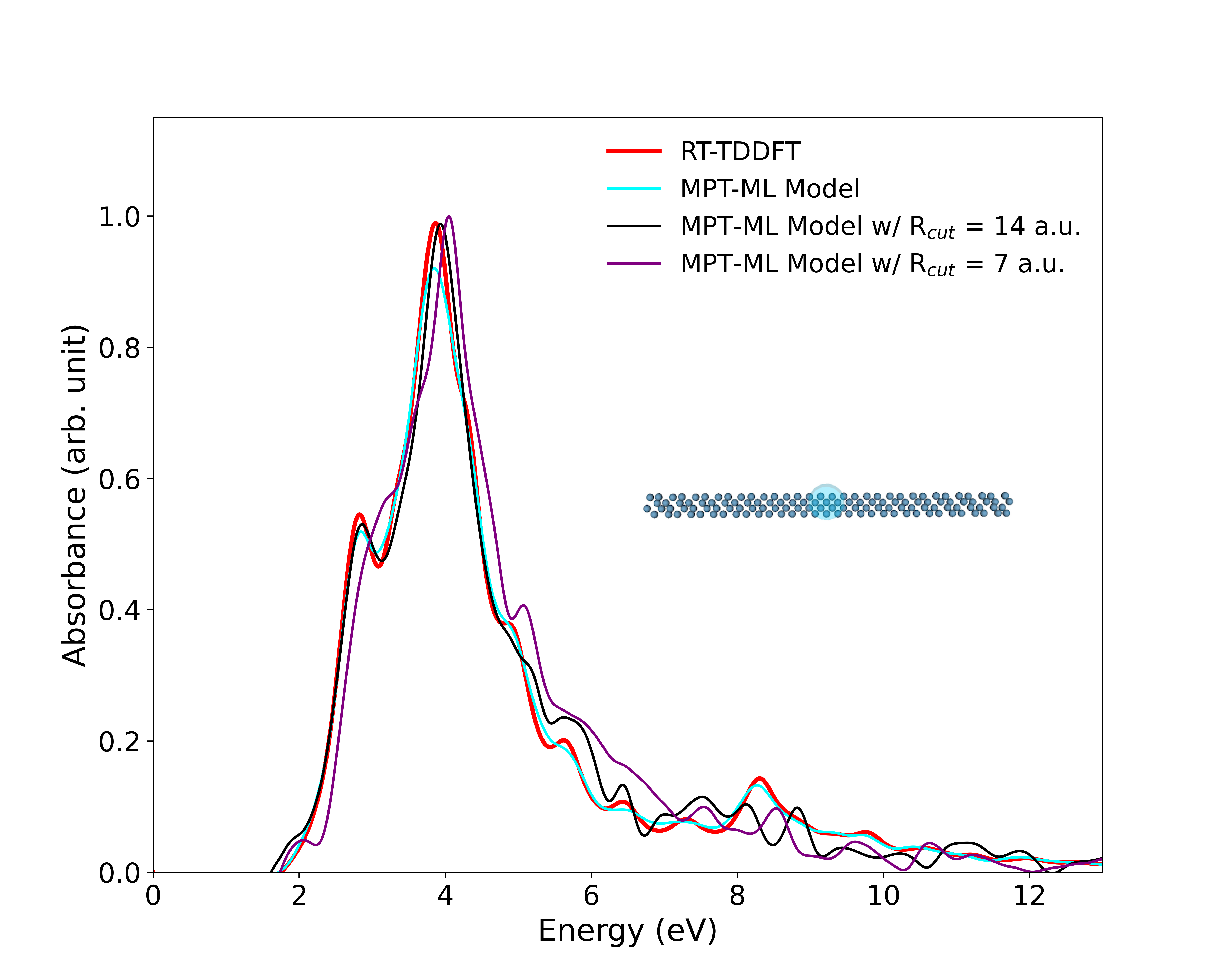}

  \caption{Optical absorption spectra of crystalline silicon, calculated with the 512-electron (128-atom) simulation cell using RT-TDDFT simulation and the MPT-ML model.
  The effect of having the varying cutoff radii ($R_{cut}$) on the MPT-ML model is shown. The MPT-ML models use the ridge regression, including the first and second order of moments. The inset shows the simulation cell used, indicating the size of 7 a.u. radius with the blue shade.
  }
  \label{fig:csi2}
\end{figure}

% Silicon
\noindent
\textbf{Crystalline Silicon:}
For modeling the optical absorption spectrum of crystalline silicon, we use an elongated supercell that consists of 128 silicon atoms, following our previous work \cite{chris_hybrid}. 
The PBE approximation was used for the XC functional, and ONCV pseudopotentials were used with a 15 Ry cutoff
for the planewave kinetic energy cutoff for the KS orbitals. The enforced time reversal symmetry (ETRS) \cite{ETRS} integrator was used to perform RT-TDDFT simulation for a total of 600 a.u. simulation time with 0.2 a.u. time steps. 
A delta kick was applied to excite the system in the direction of the elongation with the field strength of 0.001 a.u. 
Figure \ref{fig:csi} shows the spectrum obtained using the MPT-ML model along with the RT-TDDFT result. 
Unlike for the water case discussed above, including also the second-order moments does not straightforwardly improve the linear model spectrum. 
While the overall shape is improved especially the high energy region (above 5 eV), the inclusion of the second-order moments introduced an artificial peak around 1.5 eV. Here, the use of ridge regression technique for reducing the overfitting problem helps significantly, eliminating the unphysical peak below 2 eV.
Figure \ref{fig:csi2} shows how the use of the cutoff radius affect the spectrum.
While the prominent peak at 2.8 eV is largely absent with $R_{cut}=7 a.u.$, the cutoff radius of $R_{cut}=14 a.u.$ is already large enough to capture the essential features of the optical absorption spectrum here. 
As summarized in Table \ref{tbl:example}, using the cutoff radius significantly reduces the number of required parameters for the machine-learning by an order of magnitude. 

\section{Cross-validation and CPU time requirement}

\begin{figure}[H]
  \includegraphics[ width=0.9\linewidth]{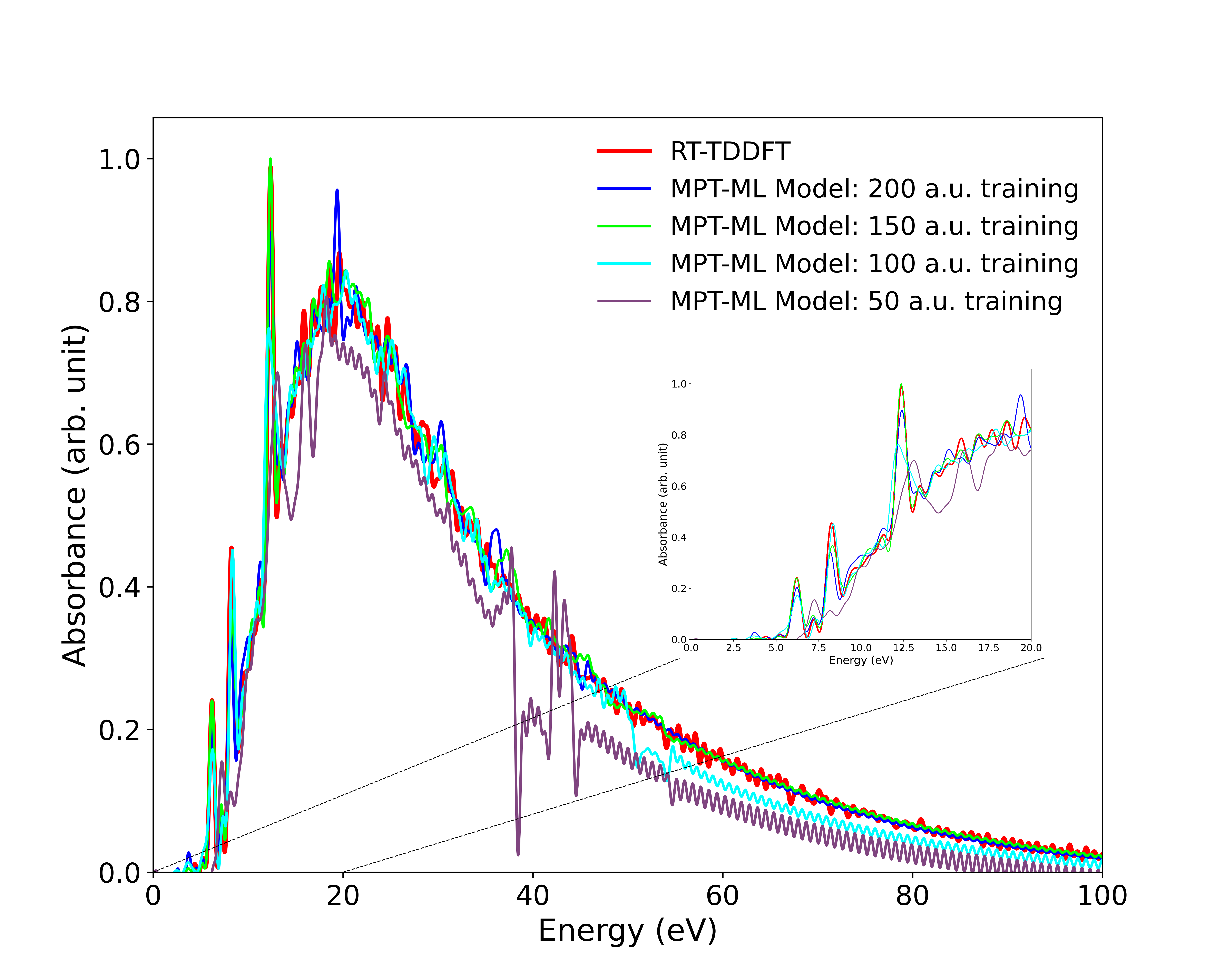}
  \caption{Cross-validation of MPT-ML models using a shorter training time. The system is a single water molecule and the MPT-ML models used up to the second order of moments. } 
  \label{fig:water_cross}
\end{figure}

We comment on the cross-validation and CPU time requirement of the MPT-ML model discussed above in this section. 
In this proof-of-principle work for the new MPT-ML model approach, our aim here was to demonstrate its efficacy by reproducing the RT-TDDFT simulation result (also the training set) using the moment propagation theory (MPT). We trained the equation-of-motion of the MPT using the machine-learning approach. 
A natural question is whether the MPT-ML model would have been able to predict the RT-TDDFT simulation result with a smaller training data set. We focus here on the single water molecule system for simplicity to answer this question, and we consider the model that includes both the first-order and second-order moments.
Figure \ref{fig:water_cross} shows how the optical absorption spectrum from the MPT-ML model changes when the training data set was obtained from RT-TDDFT simulations performed for the duration of 200, 150, 100, and 50 a.u. The reference RT-TDDFT simulation result is from the 200 a.u. RT-TDDFT simulation. 
As can be seen in Figure \ref{fig:water_cross}, the optical absorption spectrum including the prominent peaks is well reproduced already with the training data set from the shorter 100 a.u. RT-TDDFT simulation. As expected, with increasingly larger data sets, the spectrum approaches closer to that of the 200 a.u. RT-TDDFT simulation (i.e. ``RT-TDDFT" in Figure \ref{fig:water_cross}). 

Table \ref{tbl:cputime} shows the CPU time used for each part in the workflow (see Fig. \ref{fig:work}) for selected systems (a water molecule, condensed matter system consisting of 162 waters, and crystalline silicon). As can be seen, even with the additional CPU time required for training the MPT-ML model, the computational cost saving gained by using the MPT-ML model is significant; the computational time is reduced by several orders of magnitude. 
For instance, in the case of the simulation with 162 waters 
($R_{cut}=7 a.u.$), 
the CPU time required by
the MPT-ML simulation is $1.41 \times 10^4$ times lower than that of the RT-TDDFT simulation.
The computational scaling of matrix operations (such as diagonalization) required for the MPT-ML model scales with $O(n^3)$ where $n$ is the number of moments. 
This scaling can be further improved if the diagonalization (and other matrix operations) can be approximated by $m$ block diagonal matrices of equal size; this would reduce the computational scaling to $O(\frac{n^3}{m^2})$.

\begin{table}[!ht]
  \caption{Total CPU time for selected systems. Simulation time (time step) for the water molecule, the liquid water, and crystalline silicon are 200 a.u. (0.2 a.u.), 250 a.u. (0.1 a.u.), and 600 a.u. (0.2 a.u.), respectively. The simulations were performed on 88 processors on 2 Broadwell nodes (Intel Xeon E5-2699A v4-2.4 GHz) of the Dogwood Cluster at the University of North Carolina at Chapel Hill. }
  \label{tbl:cputime}
  \centering
  \begin{tabular}{llllll}
  \hline
    System & RT-TDDFT (s) & MPT-ML training (s) & MPT-ML simulation (s) \\ \hline
    Water Molecule& $6.85 \times 10^{6}$ & $2.49 \times 10^{-2}$  & $7.10 \times 10^{-2}$ \\ 
    162 Waters ($R_{cut} = 7$ a.u.) & $7.91 \times 10^{7}$ & $9.70 \times 10^{1}$ & $5.62 \times 10^{3}$ \\ 
    Silicon ($R_{cut} = 7$ a.u.) & $9.72 \times 10^{6}$ & $2.87 \times 10^{0}$ & $2.47 \times 10^{2}$ \\ \hline
\end{tabular}
\end{table}

% Conclusion
\section{Conclusions}
While TDDFT provides a particularly convenient theoretical formalism for simulating the quantum dynamics of electrons from first principles, RT-TDDFT simulation remains computationally intensive for studying many complex chemical systems \cite{doi:10.1063/5.0057587}. At the same time, data-driven modeling has become increasingly popular in many fields, especially for molecular dynamics simulation of atoms in recent years \cite{doi:10.1021/acs.chemrev.0c01111}. 
On the other hand, the electron dynamics remains as one of the challenging cases for applying data-driven approaches like ML\cite{Schiffer_ANN2}. In this work, we showed how the recently formulated MPT\cite{Boyer_2024} offers a powerful framework for machine-learning the quantum dynamics of electrons when it is combined with the RT-TDDFT simulation in the Wannier gauge\cite{10.1063/1.5095631}. MPT derives the equations of motion for all orders of moments. Due to the highly localized nature of individual MLWFs, we can anticipate that only low-order moments might be necessary for an accurate description. 
However, even for the low-order moments, their second-order time derivatives are highly complicated to calculate in practice. 
As done in the case of classical MD simulation, we applied the ML technique for approximating the second-order time derivatives by training them against the first-principles simulation \cite{10.1063/5.0155600}. 
We showed how this MPT-ML approach can be used to accurately calculate the optical absorption spectra of various systems from small gas-phase molecules to condensed phased systems even with a simple machine-learning method (i.e. linear model). For condensed matter systems, we also examined the nearsightedness principle of electrons to exploit the short-range nature of their influence to significantly reduce the number of parameters to be trained. 

This work thus far remains a proof-of-principle demonstration for real systems using first-principles calculation. At the same time, one can already realize how this MPT-ML approach can significantly benefit the field especially when using advanced XC functionals like hybrids, which are an order of magnitude computationally more expensive than standard XC functionals even with recent advancements \cite{chris_hybrid, kokott2024efficientallelectronhybriddensity}.
While this work focused on the use of the MPT-ML
approach for optical absorption spectrum, linear-response property, we envision it extended for studying more complicated non-equilibrium electron dynamics phenomena in future work.

\section{Supporting Information}
Supporting Information includes a discussion about propagation using the Taylor series expansion method and close-up views of the absorbance peaks of the molecules.

\begin{acknowledgement}

N.B. was supported by the Summer Undergraduate Research Fellowship (SURF) at the University of North Carolina at Chapel Hill. This work was supported by the National Science Foundation, under No. CHE-1954894.
\end{acknowledgement}

%_____ Author Declarations
\section*{Author Declarations}

\subsection*{Conflict of Interest}

The authors have no conflicts to disclose.

\subsection*{Author Contributions}
N.B. led the work and performed all the calculations. N.B. and Y.K. conceived of the presented idea. All authors discussed the results and contributed to the final manuscript.

\subsection*{Data Availability}
The data that support the findings of this study are available from the corresponding author upon reasonable request.

\clearpage

% end

\bibliography{SecondPaper}

\providecommand{\latin}[1]{#1}
\makeatletter
\providecommand{\doi}
  {\begingroup\let\do\@makeother\dospecials
  \catcode`\{=1 \catcode`\}=2 \doi@aux}
\providecommand{\doi@aux}[1]{\endgroup\texttt{#1}}
\makeatother
\providecommand*\mcitethebibliography{\thebibliography}
\csname @ifundefined\endcsname{endmcitethebibliography}  {\let\endmcitethebibliography\endthebibliography}{}
\begin{mcitethebibliography}{54}
\providecommand*\natexlab[1]{#1}
\providecommand*\mciteSetBstSublistMode[1]{}
\providecommand*\mciteSetBstMaxWidthForm[2]{}
\providecommand*\mciteBstWouldAddEndPuncttrue
  {\def\EndOfBibitem{\unskip.}}
\providecommand*\mciteBstWouldAddEndPunctfalse
  {\let\EndOfBibitem\relax}
\providecommand*\mciteSetBstMidEndSepPunct[3]{}
\providecommand*\mciteSetBstSublistLabelBeginEnd[3]{}
\providecommand*\EndOfBibitem{}
\mciteSetBstSublistMode{f}
\mciteSetBstMaxWidthForm{subitem}{(\alph{mcitesubitemcount})}
\mciteSetBstSublistLabelBeginEnd
  {\mcitemaxwidthsubitemform\space}
  {\relax}
  {\relax}

\bibitem[Shepard \latin{et~al.}(2021)Shepard, Zhou, Yost, Yao, and Kanai]{doi:10.1063/5.0057587}
Shepard,~C.; Zhou,~R.; Yost,~D.~C.; Yao,~Y.; Kanai,~Y. Simulating electronic excitation and dynamics with real-time propagation approach to TDDFT within plane-wave pseudopotential formulation. \emph{The Journal of Chemical Physics} \textbf{2021}, \emph{155}, 100901\relax
\mciteBstWouldAddEndPuncttrue
\mciteSetBstMidEndSepPunct{\mcitedefaultmidpunct}
{\mcitedefaultendpunct}{\mcitedefaultseppunct}\relax
\EndOfBibitem
\bibitem[Draeger \latin{et~al.}(2017)Draeger, Andrade, Gunnels, Bhatele, Schleife, and Correa]{DRAEGER2017205}
Draeger,~E.~W.; Andrade,~X.; Gunnels,~J.~A.; Bhatele,~A.; Schleife,~A.; Correa,~A.~A. Massively parallel first-principles simulation of electron dynamics in materials. \emph{Journal of Parallel and Distributed Computing} \textbf{2017}, \emph{106}, 205--214\relax
\mciteBstWouldAddEndPuncttrue
\mciteSetBstMidEndSepPunct{\mcitedefaultmidpunct}
{\mcitedefaultendpunct}{\mcitedefaultseppunct}\relax
\EndOfBibitem
\bibitem[Xu \latin{et~al.}(2024)Xu, Carney, Zhou, Shepard, and Kanai]{doi:10.1021/jacs.3c08226}
Xu,~J.; Carney,~T.~E.; Zhou,~R.; Shepard,~C.; Kanai,~Y. Real-Time Time-Dependent Density Functional Theory for Simulating Nonequilibrium Electron Dynamics. \emph{Journal of the American Chemical Society} \textbf{2024}, \emph{146}, 5011--5029, PMID: 38362887\relax
\mciteBstWouldAddEndPuncttrue
\mciteSetBstMidEndSepPunct{\mcitedefaultmidpunct}
{\mcitedefaultendpunct}{\mcitedefaultseppunct}\relax
\EndOfBibitem
\bibitem[Shepard and Kanai(2023)Shepard, and Kanai]{doi:10.1021/acs.jpcb.3c05446}
Shepard,~C.; Kanai,~Y. Ion-Type Dependence of DNA Electronic Excitation in Water under Proton, alpha-Particle, and Carbon Ion Irradiation: A First-Principles Simulation Study. \emph{The Journal of Physical Chemistry B} \textbf{2023}, \emph{127}, 10700--10709, PMID: 37943091\relax
\mciteBstWouldAddEndPuncttrue
\mciteSetBstMidEndSepPunct{\mcitedefaultmidpunct}
{\mcitedefaultendpunct}{\mcitedefaultseppunct}\relax
\EndOfBibitem
\bibitem[Sun \latin{et~al.}(2021)Sun, Lee, Kononov, Schleife, and Ullrich]{PhysRevLett.127.077401}
Sun,~J.; Lee,~C.-W.; Kononov,~A.; Schleife,~A.; Ullrich,~C.~A. Real-Time Exciton Dynamics with Time-Dependent Density-Functional Theory. \emph{Phys. Rev. Lett.} \textbf{2021}, \emph{127}, 077401\relax
\mciteBstWouldAddEndPuncttrue
\mciteSetBstMidEndSepPunct{\mcitedefaultmidpunct}
{\mcitedefaultendpunct}{\mcitedefaultseppunct}\relax
\EndOfBibitem
\bibitem[Hekele \latin{et~al.}(2021)Hekele, Yao, Kanai, Blum, and Kratzer]{10.1063/5.0066753}
Hekele,~J.; Yao,~Y.; Kanai,~Y.; Blum,~V.; Kratzer,~P. {All-electron real-time and imaginary-time time-dependent density functional theory within a numeric atom-centered basis function framework}. \emph{The Journal of Chemical Physics} \textbf{2021}, \emph{155}, 154801\relax
\mciteBstWouldAddEndPuncttrue
\mciteSetBstMidEndSepPunct{\mcitedefaultmidpunct}
{\mcitedefaultendpunct}{\mcitedefaultseppunct}\relax
\EndOfBibitem
\bibitem[Senanayake \latin{et~al.}(2019)Senanayake, Lingerfelt, Kuda-Singappulige, Li, and Aikens]{doi:10.1021/acs.jpcc.9b00296}
Senanayake,~R.~D.; Lingerfelt,~D.~B.; Kuda-Singappulige,~G.~U.; Li,~X.; Aikens,~C.~M. Real-Time TDDFT Investigation of Optical Absorption in Gold Nanowires. \emph{The Journal of Physical Chemistry C} \textbf{2019}, \emph{123}, 14734--14745\relax
\mciteBstWouldAddEndPuncttrue
\mciteSetBstMidEndSepPunct{\mcitedefaultmidpunct}
{\mcitedefaultendpunct}{\mcitedefaultseppunct}\relax
\EndOfBibitem
\bibitem[Trepl \latin{et~al.}(2022)Trepl, Schelter, and Kümmel]{doi:10.1021/acs.jctc.2c00600}
Trepl,~T.; Schelter,~I.; Kümmel,~S. Analyzing Excitation-Energy Transfer Based on the Time-Dependent Density Functional Theory in Real Time. \emph{Journal of Chemical Theory and Computation} \textbf{2022}, \emph{18}, 6577--6587, PMID: 36268773\relax
\mciteBstWouldAddEndPuncttrue
\mciteSetBstMidEndSepPunct{\mcitedefaultmidpunct}
{\mcitedefaultendpunct}{\mcitedefaultseppunct}\relax
\EndOfBibitem
\bibitem[Yang \latin{et~al.}(2020)Yang, Pei, Deng, Mao, Wu, Yang, Wang, Aikens, Liang, and Shao]{D0CP04206D}
Yang,~J.; Pei,~Z.; Deng,~J.; Mao,~Y.; Wu,~Q.; Yang,~Z.; Wang,~B.; Aikens,~C.~M.; Liang,~W.; Shao,~Y. Analysis and visualization of energy densities. I. Insights from real-time time-dependent density functional theory simulations. \emph{Phys. Chem. Chem. Phys.} \textbf{2020}, \emph{22}, 26838--26851\relax
\mciteBstWouldAddEndPuncttrue
\mciteSetBstMidEndSepPunct{\mcitedefaultmidpunct}
{\mcitedefaultendpunct}{\mcitedefaultseppunct}\relax
\EndOfBibitem
\bibitem[Herring and Montemore(2023)Herring, and Montemore]{doi:10.1021/acsnanoscienceau.2c00061}
Herring,~C.~J.; Montemore,~M.~M. Recent Advances in Real-Time Time-Dependent Density Functional Theory Simulations of Plasmonic Nanostructures and Plasmonic Photocatalysis. \emph{ACS Nanoscience Au} \textbf{2023}, \emph{3}, 269--279\relax
\mciteBstWouldAddEndPuncttrue
\mciteSetBstMidEndSepPunct{\mcitedefaultmidpunct}
{\mcitedefaultendpunct}{\mcitedefaultseppunct}\relax
\EndOfBibitem
\bibitem[Ma \latin{et~al.}(2015)Ma, Wang, and Wang]{Ma2015}
Ma,~J.; Wang,~Z.; Wang,~L.-W. Interplay between plasmon and single-particle excitations in a metal nanocluster. \emph{Nature Communications} \textbf{2015}, \emph{6}, 10107\relax
\mciteBstWouldAddEndPuncttrue
\mciteSetBstMidEndSepPunct{\mcitedefaultmidpunct}
{\mcitedefaultendpunct}{\mcitedefaultseppunct}\relax
\EndOfBibitem
\bibitem[Mantela \latin{et~al.}(2021)Mantela, Morphis, Lambropoulos, Simserides, and Di~Felice]{doi:10.1021/acs.jpcb.0c11489}
Mantela,~M.; Morphis,~A.; Lambropoulos,~K.; Simserides,~C.; Di~Felice,~R. Effects of Structural Dynamics on Charge Carrier Transfer in B-DNA: A Combined MD and RT-TDDFT Study. \emph{The Journal of Physical Chemistry B} \textbf{2021}, \emph{125}, 3986--4003, PMID: 33857373\relax
\mciteBstWouldAddEndPuncttrue
\mciteSetBstMidEndSepPunct{\mcitedefaultmidpunct}
{\mcitedefaultendpunct}{\mcitedefaultseppunct}\relax
\EndOfBibitem
\bibitem[Provorse and Isborn(2016)Provorse, and Isborn]{https://doi.org/10.1002/qua.25096}
Provorse,~M.~R.; Isborn,~C.~M. Electron dynamics with real-time time-dependent density functional theory. \emph{International Journal of Quantum Chemistry} \textbf{2016}, \emph{116}, 739--749\relax
\mciteBstWouldAddEndPuncttrue
\mciteSetBstMidEndSepPunct{\mcitedefaultmidpunct}
{\mcitedefaultendpunct}{\mcitedefaultseppunct}\relax
\EndOfBibitem
\bibitem[Makkonen \latin{et~al.}(2021)Makkonen, Rossi, Larsen, Lopez-Acevedo, Rinke, Kuisma, and Chen]{10.1063/5.0038904}
Makkonen,~E.; Rossi,~T.~P.; Larsen,~A.~H.; Lopez-Acevedo,~O.; Rinke,~P.; Kuisma,~M.; Chen,~X. {Real-time time-dependent density functional theory implementation of electronic circular dichroism applied to nanoscale metal–organic clusters}. \emph{The Journal of Chemical Physics} \textbf{2021}, \emph{154}, 114102\relax
\mciteBstWouldAddEndPuncttrue
\mciteSetBstMidEndSepPunct{\mcitedefaultmidpunct}
{\mcitedefaultendpunct}{\mcitedefaultseppunct}\relax
\EndOfBibitem
\bibitem[Rizzi \latin{et~al.}(2016)Rizzi, Todorov, Kohanoff, and Correa]{PhysRevB.93.024306}
Rizzi,~V.; Todorov,~T.~N.; Kohanoff,~J.~J.; Correa,~A.~A. Electron-phonon thermalization in a scalable method for real-time quantum dynamics. \emph{Phys. Rev. B} \textbf{2016}, \emph{93}, 024306\relax
\mciteBstWouldAddEndPuncttrue
\mciteSetBstMidEndSepPunct{\mcitedefaultmidpunct}
{\mcitedefaultendpunct}{\mcitedefaultseppunct}\relax
\EndOfBibitem
\bibitem[Woźniak and Moszyński(2024)Woźniak, and Moszyński]{doi:10.1021/acs.jpca.3c07865}
Woźniak,~A.~P.; Moszyński,~R. Modeling of High-Harmonic Generation in the C60 Fullerene Using Ab Initio, DFT-Based, and Semiempirical Methods. \emph{The Journal of Physical Chemistry A} \textbf{2024}, \emph{128}, 2683--2702, PMID: 38534023\relax
\mciteBstWouldAddEndPuncttrue
\mciteSetBstMidEndSepPunct{\mcitedefaultmidpunct}
{\mcitedefaultendpunct}{\mcitedefaultseppunct}\relax
\EndOfBibitem
\bibitem[Okyay \latin{et~al.}(2022)Okyay, Sato, Kim, Yan, Jin, and Park]{Okyay2022}
Okyay,~M.~S.; Sato,~S.~A.; Kim,~K.~W.; Yan,~B.; Jin,~H.; Park,~N. Second harmonic Hall responses of insulators as a probe of Berry curvature dipole. \emph{Communications Physics} \textbf{2022}, \emph{5}, 303\relax
\mciteBstWouldAddEndPuncttrue
\mciteSetBstMidEndSepPunct{\mcitedefaultmidpunct}
{\mcitedefaultendpunct}{\mcitedefaultseppunct}\relax
\EndOfBibitem
\bibitem[Shepard \latin{et~al.}(2023)Shepard, Yost, and Kanai]{PhysRevLett.130.118401}
Shepard,~C.; Yost,~D.~C.; Kanai,~Y. Electronic Excitation Response of DNA to High-Energy Proton Radiation in Water. \emph{Phys. Rev. Lett.} \textbf{2023}, \emph{130}, 118401\relax
\mciteBstWouldAddEndPuncttrue
\mciteSetBstMidEndSepPunct{\mcitedefaultmidpunct}
{\mcitedefaultendpunct}{\mcitedefaultseppunct}\relax
\EndOfBibitem
\bibitem[Ramakrishna \latin{et~al.}(2023)Ramakrishna, Lokamani, Baczewski, Vorberger, and Cangi]{PhysRevB.107.115131}
Ramakrishna,~K.; Lokamani,~M.; Baczewski,~A.; Vorberger,~J.; Cangi,~A. Electrical conductivity of iron in Earth's core from microscopic Ohm's law. \emph{Phys. Rev. B} \textbf{2023}, \emph{107}, 115131\relax
\mciteBstWouldAddEndPuncttrue
\mciteSetBstMidEndSepPunct{\mcitedefaultmidpunct}
{\mcitedefaultendpunct}{\mcitedefaultseppunct}\relax
\EndOfBibitem
\bibitem[Yamijala \latin{et~al.}(2022)Yamijala, Shinde, Hanasaki, Ali, and Wong]{YAMIJALA2022127026}
Yamijala,~S.~S.; Shinde,~R.; Hanasaki,~K.; Ali,~Z.~A.; Wong,~B.~M. Photo-induced degradation of PFASs: Excited-state mechanisms from real-time time-dependent density functional theory. \emph{Journal of Hazardous Materials} \textbf{2022}, \emph{423}, 127026\relax
\mciteBstWouldAddEndPuncttrue
\mciteSetBstMidEndSepPunct{\mcitedefaultmidpunct}
{\mcitedefaultendpunct}{\mcitedefaultseppunct}\relax
\EndOfBibitem
\bibitem[Moitra \latin{et~al.}(2023)Moitra, Konecny, Kadek, Rubio, and Repisky]{doi:10.1021/acs.jpclett.2c03599}
Moitra,~T.; Konecny,~L.; Kadek,~M.; Rubio,~A.; Repisky,~M. Accurate Relativistic Real-Time Time-Dependent Density Functional Theory for Valence and Core Attosecond Transient Absorption Spectroscopy. \emph{The Journal of Physical Chemistry Letters} \textbf{2023}, \emph{14}, 1714--1724, PMID: 36757216\relax
\mciteBstWouldAddEndPuncttrue
\mciteSetBstMidEndSepPunct{\mcitedefaultmidpunct}
{\mcitedefaultendpunct}{\mcitedefaultseppunct}\relax
\EndOfBibitem
\bibitem[He \latin{et~al.}(2021)He, Li, Bandyopadhyay, and Frauenheim]{doi:10.1021/acs.nanolett.1c00520}
He,~J.; Li,~S.; Bandyopadhyay,~A.; Frauenheim,~T. Unravelling Photoinduced Interlayer Spin Transfer Dynamics in Two-Dimensional Nonmagnetic-Ferromagnetic van der Waals Heterostructures. \emph{Nano Letters} \textbf{2021}, \emph{21}, 3237--3244, PMID: 33749285\relax
\mciteBstWouldAddEndPuncttrue
\mciteSetBstMidEndSepPunct{\mcitedefaultmidpunct}
{\mcitedefaultendpunct}{\mcitedefaultseppunct}\relax
\EndOfBibitem
\bibitem[Tancogne-Dejean \latin{et~al.}(2020)Tancogne-Dejean, Eich, and Rubio]{doi:10.1021/acs.jctc.9b01064}
Tancogne-Dejean,~N.; Eich,~F.~G.; Rubio,~A. Time-Dependent Magnons from First Principles. \emph{Journal of Chemical Theory and Computation} \textbf{2020}, \emph{16}, 1007--1017, PMID: 31922758\relax
\mciteBstWouldAddEndPuncttrue
\mciteSetBstMidEndSepPunct{\mcitedefaultmidpunct}
{\mcitedefaultendpunct}{\mcitedefaultseppunct}\relax
\EndOfBibitem
\bibitem[Yao \latin{et~al.}(2019)Yao, Yost, and Kanai]{PhysRevLett.123.066401}
Yao,~Y.; Yost,~D.~C.; Kanai,~Y. $K$-Shell Core-Electron Excitations in Electronic Stopping of Protons in Water from First Principles. \emph{Phys. Rev. Lett.} \textbf{2019}, \emph{123}, 066401\relax
\mciteBstWouldAddEndPuncttrue
\mciteSetBstMidEndSepPunct{\mcitedefaultmidpunct}
{\mcitedefaultendpunct}{\mcitedefaultseppunct}\relax
\EndOfBibitem
\bibitem[Peng \latin{et~al.}(2015)Peng, Lingerfelt, Ding, Aikens, and Li]{doi:10.1021/acs.jpcc.5b00263}
Peng,~B.; Lingerfelt,~D.~B.; Ding,~F.; Aikens,~C.~M.; Li,~X. Real-Time TDDFT Studies of Exciton Decay and Transfer in Silver Nanowire Arrays. \emph{The Journal of Physical Chemistry C} \textbf{2015}, \emph{119}, 6421--6427\relax
\mciteBstWouldAddEndPuncttrue
\mciteSetBstMidEndSepPunct{\mcitedefaultmidpunct}
{\mcitedefaultendpunct}{\mcitedefaultseppunct}\relax
\EndOfBibitem
\bibitem[Miyamoto \latin{et~al.}(2017)Miyamoto, Zhang, Cheng, and Rubio]{PhysRevB.96.115451}
Miyamoto,~Y.; Zhang,~H.; Cheng,~X.; Rubio,~A. Modeling of laser-pulse induced water decomposition on two-dimensional materials by simulations based on time-dependent density functional theory. \emph{Phys. Rev. B} \textbf{2017}, \emph{96}, 115451\relax
\mciteBstWouldAddEndPuncttrue
\mciteSetBstMidEndSepPunct{\mcitedefaultmidpunct}
{\mcitedefaultendpunct}{\mcitedefaultseppunct}\relax
\EndOfBibitem
\bibitem[Jia and Lin(2019)Jia, and Lin]{JIA201921}
Jia,~W.; Lin,~L. Fast real-time time-dependent hybrid functional calculations with the parallel transport gauge and the adaptively compressed exchange formulation. \emph{Computer Physics Communications} \textbf{2019}, \emph{240}, 21--29\relax
\mciteBstWouldAddEndPuncttrue
\mciteSetBstMidEndSepPunct{\mcitedefaultmidpunct}
{\mcitedefaultendpunct}{\mcitedefaultseppunct}\relax
\EndOfBibitem
\bibitem[Yost \latin{et~al.}(2019)Yost, Yao, and Kanai]{10.1063/1.5095631}
Yost,~D.~C.; Yao,~Y.; Kanai,~Y. {Propagation of maximally localized Wannier functions in real-time TDDFT}. \emph{The Journal of Chemical Physics} \textbf{2019}, \emph{150}, 194113\relax
\mciteBstWouldAddEndPuncttrue
\mciteSetBstMidEndSepPunct{\mcitedefaultmidpunct}
{\mcitedefaultendpunct}{\mcitedefaultseppunct}\relax
\EndOfBibitem
\bibitem[Zhou \latin{et~al.}(2021)Zhou, Yost, and Kanai]{doi:10.1021/acs.jpclett.1c01037}
Zhou,~R.; Yost,~D.~C.; Kanai,~Y. First-Principles Demonstration of Nonadiabatic Thouless Pumping of Electrons in a Molecular System. \emph{The Journal of Physical Chemistry Letters} \textbf{2021}, \emph{12}, 4496--4503, PMID: 33956458\relax
\mciteBstWouldAddEndPuncttrue
\mciteSetBstMidEndSepPunct{\mcitedefaultmidpunct}
{\mcitedefaultendpunct}{\mcitedefaultseppunct}\relax
\EndOfBibitem
\bibitem[Shepard \latin{et~al.}(2024)Shepard, Zhou, Bost, Carney, Yao, and Kanai]{chris_hybrid}
Shepard,~C.; Zhou,~R.; Bost,~J.; Carney,~T.~E.; Yao,~Y.; Kanai,~Y. {Efficient exact exchange using Wannier functions and other related developments in planewave-pseudopotential implementation of RT-TDDFT}. \emph{The Journal of Chemical Physics} \textbf{2024}, \emph{161}, 024111\relax
\mciteBstWouldAddEndPuncttrue
\mciteSetBstMidEndSepPunct{\mcitedefaultmidpunct}
{\mcitedefaultendpunct}{\mcitedefaultseppunct}\relax
\EndOfBibitem
\bibitem[Westermayr \latin{et~al.}(2021)Westermayr, Gastegger, Schütt, and Maurer]{10.1063/5.0047760}
Westermayr,~J.; Gastegger,~M.; Schütt,~K.~T.; Maurer,~R.~J. {Perspective on integrating machine learning into computational chemistry and materials science}. \emph{The Journal of Chemical Physics} \textbf{2021}, \emph{154}, 230903\relax
\mciteBstWouldAddEndPuncttrue
\mciteSetBstMidEndSepPunct{\mcitedefaultmidpunct}
{\mcitedefaultendpunct}{\mcitedefaultseppunct}\relax
\EndOfBibitem
\bibitem[Doerr \latin{et~al.}(2021)Doerr, Majewski, Pérez, Krämer, Clementi, Noe, Giorgino, and De~Fabritiis]{doi:10.1021/acs.jctc.0c01343}
Doerr,~S.; Majewski,~M.; Pérez,~A.; Krämer,~A.; Clementi,~C.; Noe,~F.; Giorgino,~T.; De~Fabritiis,~G. TorchMD: A Deep Learning Framework for Molecular Simulations. \emph{Journal of Chemical Theory and Computation} \textbf{2021}, \emph{17}, 2355--2363, PMID: 33729795\relax
\mciteBstWouldAddEndPuncttrue
\mciteSetBstMidEndSepPunct{\mcitedefaultmidpunct}
{\mcitedefaultendpunct}{\mcitedefaultseppunct}\relax
\EndOfBibitem
\bibitem[Han \latin{et~al.}(2018)Han, Zhang, Car, and E]{Han_2018}
Han,~J.; Zhang,~L.; Car,~R.; E,~W. Deep Potential: A General Representation of a Many-Body Potential Energy Surface. \emph{Communications in Computational Physics} \textbf{2018}, \emph{23}\relax
\mciteBstWouldAddEndPuncttrue
\mciteSetBstMidEndSepPunct{\mcitedefaultmidpunct}
{\mcitedefaultendpunct}{\mcitedefaultseppunct}\relax
\EndOfBibitem
\bibitem[Secor \latin{et~al.}(2021)Secor, Soudackov, and Hammes-Schiffer]{Schiffer_ANN1}
Secor,~M.; Soudackov,~A.~V.; Hammes-Schiffer,~S. Artificial Neural Networks as Mappings between Proton Potentials, Wave Functions, Densities, and Energy Levels. \emph{The Journal of Physical Chemistry Letters} \textbf{2021}, \emph{12}, 2206--2212\relax
\mciteBstWouldAddEndPuncttrue
\mciteSetBstMidEndSepPunct{\mcitedefaultmidpunct}
{\mcitedefaultendpunct}{\mcitedefaultseppunct}\relax
\EndOfBibitem
\bibitem[Secor \latin{et~al.}(2021)Secor, Soudackov, and Hammes-Schiffer]{Schiffer_ANN2}
Secor,~M.; Soudackov,~A.~V.; Hammes-Schiffer,~S. Artificial Neural Networks as Propagators in Quantum Dynamics. \emph{The Journal of Physical Chemistry Letters} \textbf{2021}, \emph{12}, 10654--10662\relax
\mciteBstWouldAddEndPuncttrue
\mciteSetBstMidEndSepPunct{\mcitedefaultmidpunct}
{\mcitedefaultendpunct}{\mcitedefaultseppunct}\relax
\EndOfBibitem
\bibitem[Boyer \latin{et~al.}(2024)Boyer, Shepard, Zhou, Xu, and Kanai]{Boyer_2024}
Boyer,~N.~J.; Shepard,~C.; Zhou,~R.; Xu,~J.; Kanai,~Y. Theory of moment propagation for quantum dynamics in single-particle description. \emph{The Journal of Chemical Physics} \textbf{2024}, \emph{160}\relax
\mciteBstWouldAddEndPuncttrue
\mciteSetBstMidEndSepPunct{\mcitedefaultmidpunct}
{\mcitedefaultendpunct}{\mcitedefaultseppunct}\relax
\EndOfBibitem
\bibitem[Bull-Vulpe \latin{et~al.}(2023)Bull-Vulpe, Riera, Bore, and Paesani]{doi:10.1021/acs.jctc.2c00645}
Bull-Vulpe,~E.~F.; Riera,~M.; Bore,~S.~L.; Paesani,~F. Data-Driven Many-Body Potential Energy Functions for Generic Molecules: Linear Alkanes as a Proof-of-Concept Application. \emph{Journal of Chemical Theory and Computation} \textbf{2023}, \emph{19}, 4494--4509, PMID: 36113028\relax
\mciteBstWouldAddEndPuncttrue
\mciteSetBstMidEndSepPunct{\mcitedefaultmidpunct}
{\mcitedefaultendpunct}{\mcitedefaultseppunct}\relax
\EndOfBibitem
\bibitem[Hauge \latin{et~al.}(2023)Hauge, Kristiansen, Konecny, Kadek, Repisky, and Pedersen]{doi:10.1021/acs.jctc.3c00727}
Hauge,~E.; Kristiansen,~H.~E.; Konecny,~L.; Kadek,~M.; Repisky,~M.; Pedersen,~T.~B. Cost-Efficient High-Resolution Linear Absorption Spectra through Extrapolating the Dipole Moment from Real-Time Time-Dependent Electronic-Structure Theory. \emph{Journal of Chemical Theory and Computation} \textbf{2023}, \emph{19}, 7764--7775, PMID: 37874968\relax
\mciteBstWouldAddEndPuncttrue
\mciteSetBstMidEndSepPunct{\mcitedefaultmidpunct}
{\mcitedefaultendpunct}{\mcitedefaultseppunct}\relax
\EndOfBibitem
\bibitem[Resta(1998)]{PhysRevLett.80.1800}
Resta,~R. Quantum-Mechanical Position Operator in Extended Systems. \emph{Phys. Rev. Lett.} \textbf{1998}, \emph{80}, 1800--1803\relax
\mciteBstWouldAddEndPuncttrue
\mciteSetBstMidEndSepPunct{\mcitedefaultmidpunct}
{\mcitedefaultendpunct}{\mcitedefaultseppunct}\relax
\EndOfBibitem
\bibitem[Wan(2015)]{thesismoments}
Wan,~Q. First Princliples Simulations of Vibrational Spectra of Aqueous Systems. PhD thesis, University of Chicago, 2015; Available at \url{https://galligroup.uchicago.edu/Research/Quan_Wan_thesis.pdf}\relax
\mciteBstWouldAddEndPuncttrue
\mciteSetBstMidEndSepPunct{\mcitedefaultmidpunct}
{\mcitedefaultendpunct}{\mcitedefaultseppunct}\relax
\EndOfBibitem
\bibitem[Kohn(1996)]{PhysRevLett.76.3168}
Kohn,~W. Density Functional and Density Matrix Method Scaling Linearly with the Number of Atoms. \emph{Phys. Rev. Lett.} \textbf{1996}, \emph{76}, 3168--3171\relax
\mciteBstWouldAddEndPuncttrue
\mciteSetBstMidEndSepPunct{\mcitedefaultmidpunct}
{\mcitedefaultendpunct}{\mcitedefaultseppunct}\relax
\EndOfBibitem
\bibitem[Prodan and Kohn(2005)Prodan, and Kohn]{Nearsight_2}
Prodan,~E.; Kohn,~W. Nearsightedness of electronic matter. \emph{Proceedings of the National Academy of Sciences} \textbf{2005}, \emph{102}, 11635--11638\relax
\mciteBstWouldAddEndPuncttrue
\mciteSetBstMidEndSepPunct{\mcitedefaultmidpunct}
{\mcitedefaultendpunct}{\mcitedefaultseppunct}\relax
\EndOfBibitem
\bibitem[Perdew \latin{et~al.}(1996)Perdew, Burke, and Ernzerhof]{PBE}
Perdew,~J.~P.; Burke,~K.; Ernzerhof,~M. Generalized Gradient Approximation Made Simple. \emph{Phys. Rev. Lett.} \textbf{1996}, \emph{77}, 3865--3868\relax
\mciteBstWouldAddEndPuncttrue
\mciteSetBstMidEndSepPunct{\mcitedefaultmidpunct}
{\mcitedefaultendpunct}{\mcitedefaultseppunct}\relax
\EndOfBibitem
\bibitem[Schlipf and Gygi(2015)Schlipf, and Gygi]{SCHLIPF201536}
Schlipf,~M.; Gygi,~F. Optimization algorithm for the generation of ONCV pseudopotentials. \emph{Computer Physics Communications} \textbf{2015}, \emph{196}, 36--44\relax
\mciteBstWouldAddEndPuncttrue
\mciteSetBstMidEndSepPunct{\mcitedefaultmidpunct}
{\mcitedefaultendpunct}{\mcitedefaultseppunct}\relax
\EndOfBibitem
\bibitem[Castro \latin{et~al.}(2004)Castro, Marques, and Rubio]{ETRS}
Castro,~A.; Marques,~M. A.~L.; Rubio,~A. {Propagators for the time-dependent Kohn–Sham equations}. \emph{The Journal of Chemical Physics} \textbf{2004}, \emph{121}, 3425--3433\relax
\mciteBstWouldAddEndPuncttrue
\mciteSetBstMidEndSepPunct{\mcitedefaultmidpunct}
{\mcitedefaultendpunct}{\mcitedefaultseppunct}\relax
\EndOfBibitem
\bibitem[Berendsen \latin{et~al.}(1987)Berendsen, Grigera, and Straatsma]{SPC-E}
Berendsen,~H. J.~C.; Grigera,~J.~R.; Straatsma,~T.~P. The missing term in effective pair potentials. \emph{The Journal of Physical Chemistry} \textbf{1987}, \emph{91}, 6269--6271\relax
\mciteBstWouldAddEndPuncttrue
\mciteSetBstMidEndSepPunct{\mcitedefaultmidpunct}
{\mcitedefaultendpunct}{\mcitedefaultseppunct}\relax
\EndOfBibitem
\bibitem[Vanderbilt(1985)]{PhysRevB.32.8412}
Vanderbilt,~D. Optimally smooth norm-conserving pseudopotentials. \emph{Phys. Rev. B} \textbf{1985}, \emph{32}, 8412--8415\relax
\mciteBstWouldAddEndPuncttrue
\mciteSetBstMidEndSepPunct{\mcitedefaultmidpunct}
{\mcitedefaultendpunct}{\mcitedefaultseppunct}\relax
\EndOfBibitem
\bibitem[Prendergast \latin{et~al.}(2005)Prendergast, Grossman, and Galli]{water2}
Prendergast,~D.; Grossman,~J.~C.; Galli,~G. {The electronic structure of liquid water within density-functional theory}. \emph{The Journal of Chemical Physics} \textbf{2005}, \emph{123}, 014501\relax
\mciteBstWouldAddEndPuncttrue
\mciteSetBstMidEndSepPunct{\mcitedefaultmidpunct}
{\mcitedefaultendpunct}{\mcitedefaultseppunct}\relax
\EndOfBibitem
\bibitem[Shepard and Kanai(2022)Shepard, and Kanai]{water1}
Shepard,~C.; Kanai,~Y. Nonlinear electronic excitation in water under proton irradiation: a first principles study. \emph{Phys. Chem. Chem. Phys.} \textbf{2022}, \emph{24}, 5598--5603\relax
\mciteBstWouldAddEndPuncttrue
\mciteSetBstMidEndSepPunct{\mcitedefaultmidpunct}
{\mcitedefaultendpunct}{\mcitedefaultseppunct}\relax
\EndOfBibitem
\bibitem[Prodan and Kohn(2005)Prodan, and Kohn]{doi:10.1073/pnas.0505436102}
Prodan,~E.; Kohn,~W. Nearsightedness of electronic matter. \emph{Proceedings of the National Academy of Sciences} \textbf{2005}, \emph{102}, 11635--11638\relax
\mciteBstWouldAddEndPuncttrue
\mciteSetBstMidEndSepPunct{\mcitedefaultmidpunct}
{\mcitedefaultendpunct}{\mcitedefaultseppunct}\relax
\EndOfBibitem
\bibitem[Unke \latin{et~al.}(2021)Unke, Chmiela, Sauceda, Gastegger, Poltavsky, Schütt, Tkatchenko, and Müller]{doi:10.1021/acs.chemrev.0c01111}
Unke,~O.~T.; Chmiela,~S.; Sauceda,~H.~E.; Gastegger,~M.; Poltavsky,~I.; Schütt,~K.~T.; Tkatchenko,~A.; Müller,~K.-R. Machine Learning Force Fields. \emph{Chemical Reviews} \textbf{2021}, \emph{121}, 10142--10186, PMID: 33705118\relax
\mciteBstWouldAddEndPuncttrue
\mciteSetBstMidEndSepPunct{\mcitedefaultmidpunct}
{\mcitedefaultendpunct}{\mcitedefaultseppunct}\relax
\EndOfBibitem
\bibitem[Zeng \latin{et~al.}(2023)Zeng, Zhang, Lu, Mo, Li, Chen, Rynik, Huang, Li, Shi, Wang, Ye, Tuo, Yang, Ding, Li, Tisi, Zeng, Bao, Xia, Huang, Muraoka, Wang, Chang, Yuan, Bore, Cai, Lin, Wang, Xu, Zhu, Luo, Zhang, Goodall, Liang, Singh, Yao, Zhang, Wentzcovitch, Han, Liu, Jia, York, E, Car, Zhang, and Wang]{10.1063/5.0155600}
Zeng,~J. \latin{et~al.}  {DeePMD-kit v2: A software package for deep potential models}. \emph{The Journal of Chemical Physics} \textbf{2023}, \emph{159}, 054801\relax
\mciteBstWouldAddEndPuncttrue
\mciteSetBstMidEndSepPunct{\mcitedefaultmidpunct}
{\mcitedefaultendpunct}{\mcitedefaultseppunct}\relax
\EndOfBibitem
\bibitem[Kokott \latin{et~al.}(2024)Kokott, Merz, Yao, Carbogno, Rossi, Havu, Rampp, Scheffler, and Blum]{kokott2024efficientallelectronhybriddensity}
Kokott,~S.; Merz,~F.; Yao,~Y.; Carbogno,~C.; Rossi,~M.; Havu,~V.; Rampp,~M.; Scheffler,~M.; Blum,~V. Efficient All-electron Hybrid Density Functionals for Atomistic Simulations Beyond 10,000 Atoms. \emph{arXiv} \textbf{2024}, \relax
\mciteBstWouldAddEndPunctfalse
\mciteSetBstMidEndSepPunct{\mcitedefaultmidpunct}
{}{\mcitedefaultseppunct}\relax
\EndOfBibitem
\end{mcitethebibliography}

\end{document}